\documentclass[structabstract]{aa}
\usepackage{graphicx}
\usepackage{txfonts}
\usepackage{natbib}
\begin{document}
   \title{Star complexes and stellar populations in NGC 6822}

 \subtitle{Comparison with the Magellanic Clouds}

   \author{A. Karampelas
          \inst{1}
          \and
           A. Dapergolas
          \inst{2}
          \and
           E. Kontizas
          \inst{2}
          \and
           E. Livanou
          \inst{1}
          \and
           M. Kontizas
          \inst{1}
          \and
           I. Bellas-Velidis
          \inst{2}
          \and
           J. M. V\'{\i}lchez
          \inst{3}
          }

   \institute{Department of Astrophysics, Astronomy \& Mechanics,
Faculty of Physics, University of Athens, GR-15783 Athens, Greece\\
              \email{ankaramp@phys.uoa.gr}
         \and
             Institute for Astronomy and Astrophysics, National
Observatory of Athens, P.O. Box 20048, GR-11810 Athens, Greece
         \and
             Instituto de Astrof\'{\i}sica de Andaluc\'{\i}a (CSIC),
Apartado 3004, 18080 Granada, Spain
             }

   \date{Received ...; Accepted ...}


  \abstract
   {}
   {The star complexes (large scale star forming regions) of NGC 6822 were traced and mapped
   and their size distribution was compared with the size distribution of star complexes
   in the Magellanic Clouds (MCs). Furthermore, the spatial distributions of different
   age stellar populations were compared with each other.}
   {The star complexes of NGC 6822 were determined by using the isopleths, based on star
   counts, of the young stars of the galaxy, using a statistical cutoff limit in density.
   In order to map them and determine their geometrical properties, an ellipse was fitted
   to every distinct region satisfying this minimum limit. The Kolmogorov-Smirnov statistical test
   was used to study possible patterns in their size distribution. Isopleths were also used to study
   the stellar populations of NGC 6822.}
   {The star complexes of NGC 6822 were detected and a list of their positions and sizes was produced.
   Indications of hierarchical star formation, in terms of spatial
   distribution, time evolution and preferable sizes were found in NGC 6822 and the MCs.
   The spatial distribution of the various age stellar populations has indicated traces of an
   interaction in NGC 6822, dated before 350 $\pm$ 50 Myr.}
   {}


   \maketitle
%

\section{Introduction}

Stars are considered to be born inside stellar groupings, the lowest
in size being OB associations. Bigger groupings, containing star
clusters, OB associations, HII regions and individual bright stars,
are star complexes, with an age of up to $\sim$ 100 Myr (Efremov
1978, 1979). Star complexes seem to be part of a continuous star
formation hierarchy, in which stars form in hierarchically clustered
systems from sub-parcec to kiloparcec scales, following the
hierarchical distribution of the gas (Elmegreen \& Efremov 1996).
The study of the positions and ages of Cepheid variables in the
Large Magellanic Cloud (Elmegreen \& Efremov 1996) and star clusters
of the same galaxy (Efremov \& Elmegreen 1998), results in that star
formation is not only hierarchical in space, but in time as well.
Larger star forming regions have higher average ages than smaller
regions, forming stars over a longer period. The above suggestions
(hereafter Efremov \& Elmegreen model) are confirmed in subsequent
studies of the Large Magellanic Cloud (Maragoudaki et al. 1998,
Livanou et al. 2006) and the Small Magellanic Cloud (Maragoudaki et
al. 2001, Livanou et al. 2007). According to their sizes, star
complexes are empirically divided (Efremov 1987, Maragoudaki et al.
1998, Livanou et al. 2007) into stellar aggregates, stellar
complexes and stellar supercomplexes, with sizes from 150 pc to 300
pc, 300 pc to 1 kpc and greater than 1 kpc, respectively. Their
properties seem to be universal, revealing their importance in the
studies of star formation in galaxies and galaxy evolution.

NGC 6822 is a Local Group dwarf irregular galaxy, discovered by
Barnard in 1884. It is located in a relatively isolated position of
the Local Group of Galaxies, being a member of the so-called "Local
Cloud" of dwarf Irregulars (Mateo 1998), at a distance of 500 kpc.
The general properties of NGC 6822 are given in Table
\ref{properties}.

The stellar content of NGC 6822 forms a stellar halo extended more
than one degree, with the young population in a bar-like structure
(P.A. $\sim$ 10\degr, Hodge 1977) and the old population
elliptically distributed (Demers et al 2006). The Star Formation
Rate (SFR) of NGC 6822 is supposed to have been relative low and
constant during the past, but increased recently, especially in the
bar. Gallart et al. (1996b) argue for an increasing SFR during the
last 400 Myr in comparison to the past, which is more obvious for
the last 100-200 Myr, mainly in the bar, while Wyder (2001) detects
the beginning of this increase back to 600 Myr. Furthermore, Hodge
(1980) has discovered an increased star cluster Formation Rate
during the last 75-100 Myr and Skillman (1989) traced a low value of
the N/O ratio, which is characteristic of low metallicity, active
star forming dwarf galaxies.

Almost vertical to the elliptical structure of the old stellar
population lies the highly disturbed elongated disk of atomic
hydrogen (P.A. $\sim$ 130\degr, Weldrake et al. 2003). The most
striking features of the gas disk are the 1 kpc-size Giant Hole
(Hodge et al. 1991; de Blok, \& Walter 2000), the SE tidal Arm and
the NW Cloud (de Blok, \& Walter 2000). An upper limit of 130 Myr is
set for the kinematical age of the Giant Hole, while the same
quantity is estimated for the tidal Arm to be around 140 Myr (de
Blok, \& Walter 2003, 2006).

The NW Cloud is considered by de Blok, \& Walter (2000; 2006) to be
a companion galaxy of NGC 6822. This "companion" galaxy together
with NGC 6822 are involved by the same authors in an interaction
scenario during the last 300 Myr, giving a possible explanation to
the recent Star Formation excess described above, since a strong
interaction is unlikely due to the isolation of NGC 6822. The same
argument is used by Weldrake et al. (2003) against a cuspy Dark
Matter halo in NGC 6822. However, Valenzuela et al. (2007) argue
that the rotation curve of this galaxy is consistent with a cuspy
Dark Matter halo.

On the other hand, the vertical arrangement of the old stellar
population of NGC 6822 respecting to the gas disk and their
different kinematical properties, lead to the possibility of NGC
6822 being a Polar Ring galaxy (Demers et al. 2006). For the
formation of Polar Rings strong interactions are required. The
strong interaction necessary for NGC 6822 to become a Polar Ring
galaxy is placed by the same authors well before 500 Myr.

The structure of this paper is as follows: Section 2 provides a
description of the data we used. Section 3 contains the steps for
the detection of star complexes of NGC 6822, while Section 4 deals
with the size distribution of these complexes and provides a
comparison with the Magellanic Clouds. Section 5 refers to the
spatial distribution of the different age stellar populations of NGC
6822. Discussion and Conclusions are in Sections 6 and 7,
respectively.

\begin{table}
\caption{General properties of NGC 6822.} \label{properties}
\centering
\begin{tabular}{c c}
\hline\hline
  Type & $IB(s)m ^{\mathrm{a}}$ \\
  R.A. (center)      & $19^{h} 44^{m} 56.6^{s}$ \\
  Dec. (center)     & $-14^{\circ} 47^{\prime} 21^{\prime\prime}$ $^{\mathrm{a}}$ \\
  Dimensions   & $15.5^{\prime}\times13.5^{\prime}$ $^{\mathrm{a}}$ \\
  Distance    & $500$ kpc $^{\mathrm{b}}$ \\
  Distance modulus    & $23.49\pm0.05$ mag $^{\mathrm{c}}$ \\
  $M_{V}$    & $-16$ mag $^{\mathrm{b}}$ \\
  E(B-V)    & $0.25\pm0.02$ mag  $^{\mathrm{d}}$ \\
  Z    & $0.004 ^{\mathrm{e}}$ \\
  SFR    & $0.06 M_{\odot}\cdot  yr^{-1}$ $^{\mathrm{f}}$ \\
\hline
\end{tabular}
\begin{list}{}{}
\item[$^{\mathrm{a}}$] NASA/IPAC Extragalactic Database
\item[$^{\mathrm{b}}$] van den Bergh 2000
\item[$^{\mathrm{c}}$] Gallart et al. 1996a
\item[$^{\mathrm{d}}$] Massey et al. 2007
\item[$^{\mathrm{e}}$] Skillman 1989
\item[$^{\mathrm{f}}$] Mateo 1998, Israel et al. 1996
\end{list}
\end{table}

\section{Data}

\begin{figure}
\centering
\includegraphics[width=0.49\textwidth]{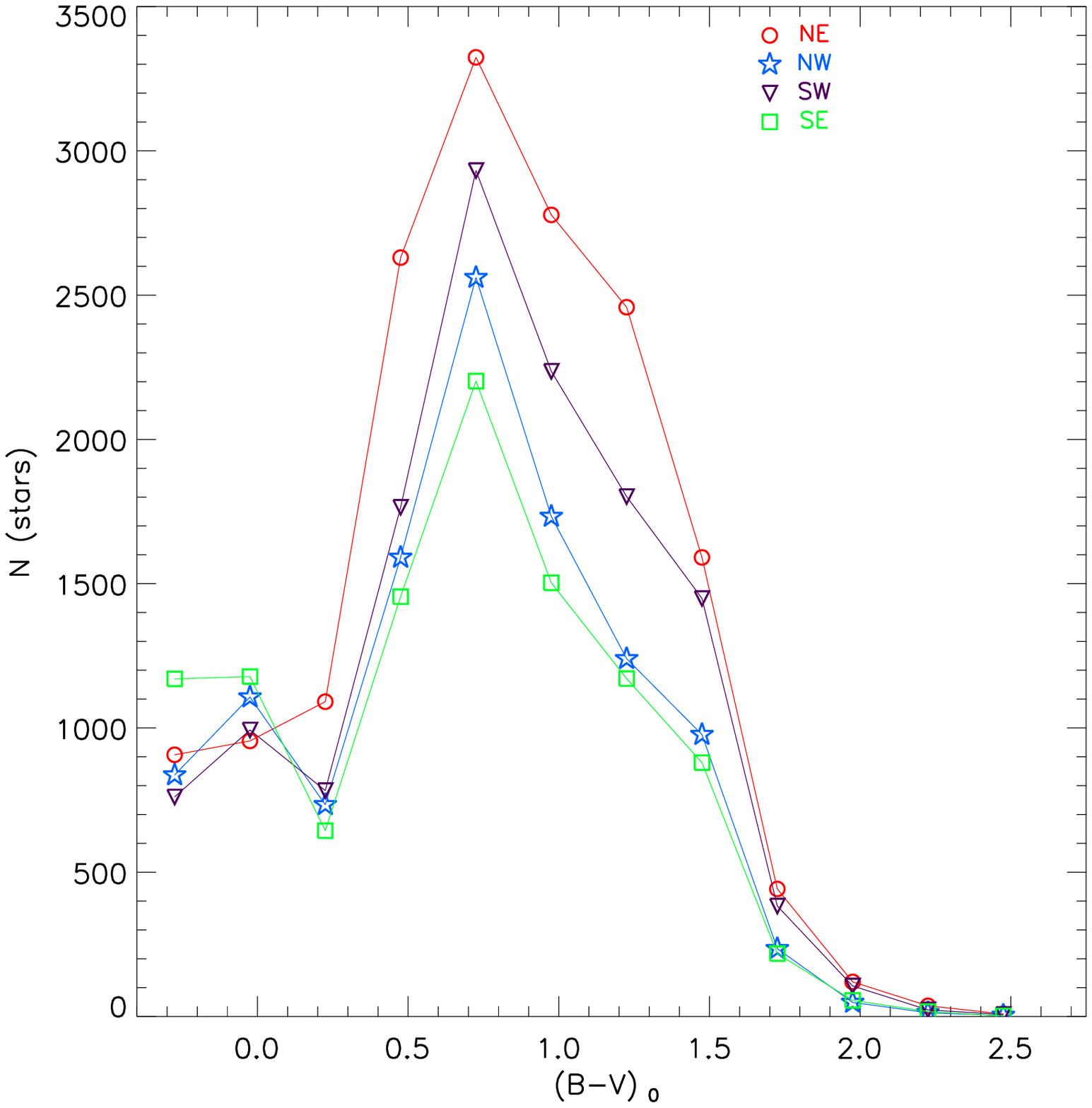}
\includegraphics[width=0.49\textwidth]{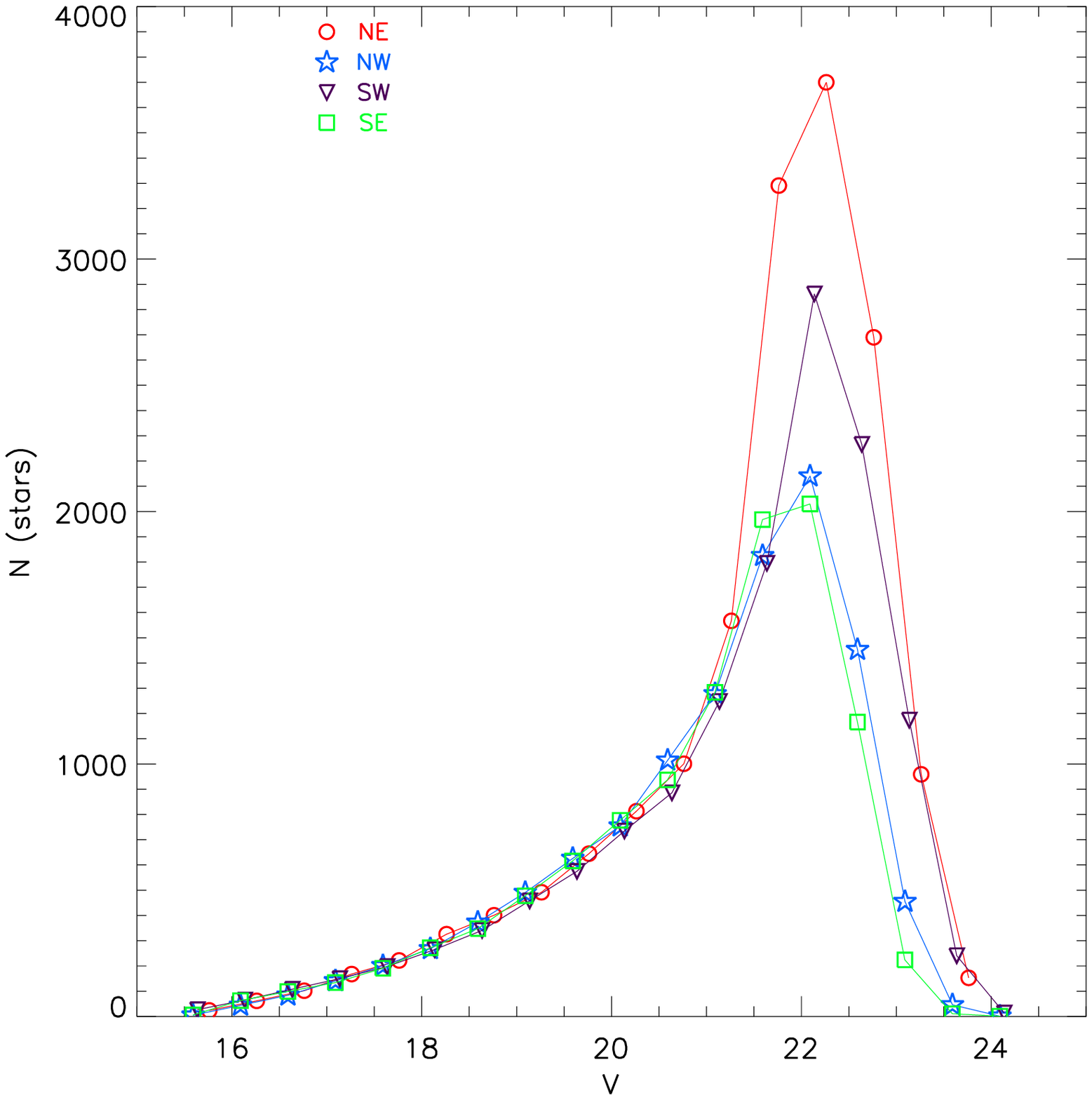}
\caption{\textbf{Upper:} Distribution of the colour of the stars
according to their relative location in the camera. \textbf{Lower:}
Luminosity function of the stars according to their relative
location in the camera.} \label{profiles}
\end{figure}

We used the NGC 6822 stellar catalogue from the "Local Group Survey"
project (Massey et al. 2007), available on the web
(ftp://ftp.lowell.edu/pub/massey/lgsurvey), based on observations
using the 4-m Blanco telescope (CTIO). The catalogue includes
position and UBVRI magnitudes of 51877 stars observed in, at least,
the B, V and R bands. The camera consists of eight $2048 \times
4096$ $pixel^{2}$ CCDs, forming a $8192 \times 8192$ $pixel^{2}$
mosaic. The Field of View is about $35^{\prime}\times35^{\prime}$,
while the scale is $0.27^{\prime\prime}$ pixel$^{-1}$. Consequently,
the stellar catalogue covers the vast majority of the large scale
structures of NGC 6822.

However, exposures with different Point-Spread Function (PSF) and
chip-to-chip PSF variations and color responses (Massey et al.
2006), may give an erroneous spatial distribution. Therefore, we
divided the catalogue of stars in four subgroups according to their
relative location in the camera and produced their distribution both
in colour and magnitude (Figure \ref{profiles}). Indeed, it appears
that the detection of red and faint stars is not homogeneous across
the field of the galaxy, being more efficient in the NE and SW areas
of the camera. This has to be taken into account where comparing the
spatial distribution of different stellar populations. On the
contrary, there is not such an effect in the blue and bright stars
used to determine the star complexes.

The CMD diagram of NGC 6822 is shown in Figure \ref{cmd_ellipses}.
Thirty stars were randomly selected from the stellar catalogue to
illustrate the photometric uncertainties. For the majority of the
stars the uncertainties are very low. However, there are
considerable uncertainties in the determination of magnitude and
colours of some stars, especially in the faint part of the CMD (see
also Table 14 in Massey et al. 2007). Finally, the luminosity
function of the stars according to their relative location in the
camera (Figure \ref{profiles}) provides an estimation of the
completeness of the data. The stellar catalogue is quite complete
until $V = 22$.

\section{Determination of Star Complexes}

\begin{figure*}[h]
   \centering
   \includegraphics[width=.6\textwidth]{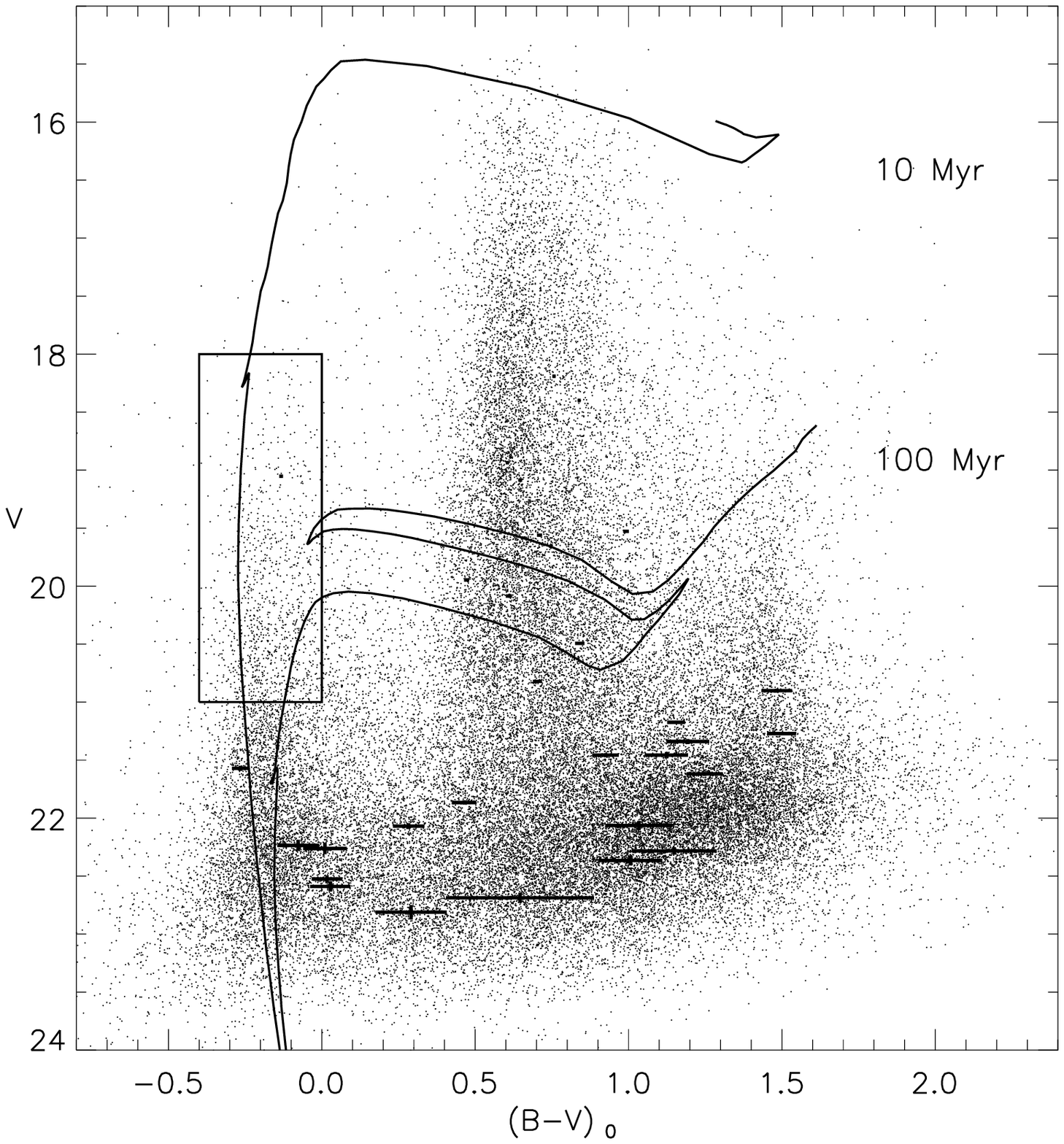}
   \includegraphics[width=.6\textwidth]{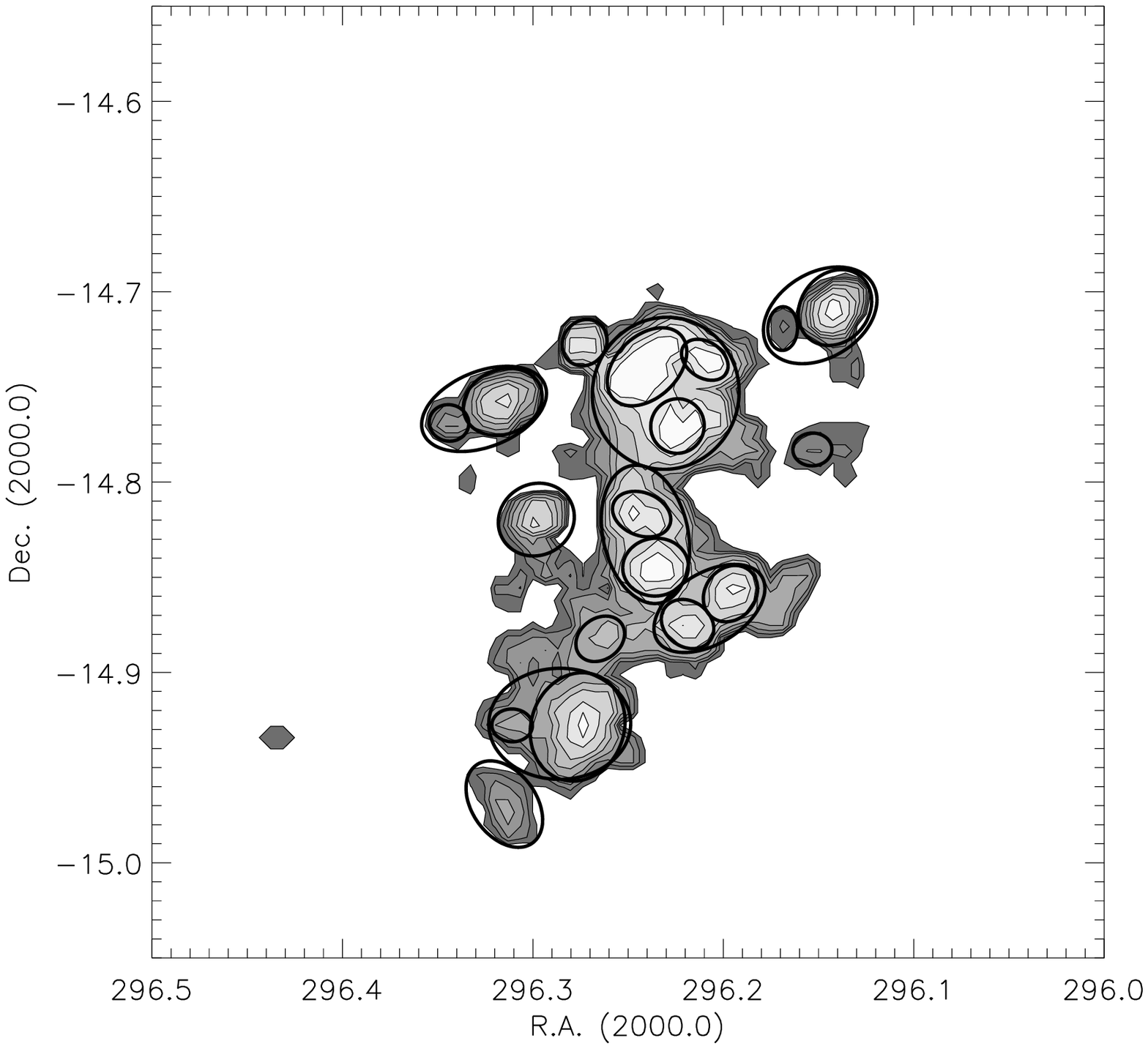}
      \caption{\textbf{Upper:} CMD diagram of NGC 6822 and photometric uncertainties in
      colour and magnitude of 30 stars randomly selected from the stellar catalogue.
      The rectangle includes the young stellar population that was used for the detection
      of star complexes, and its characteristic theoretical isochrones. \textbf{Lower:} Detected star
complexes of NGC 6822 (ellipses) plotted over the isopleths of the
surface stellar density map of the young stars.
              }
         \label{cmd_ellipses}
   \end{figure*}

The Colour-Magnitude Diagram (CMD) of NGC 6822 was constructed using
the "Local Group Survey" stellar catalogue and adopting the values
of 23.49 mag for the distance modulus (Gallart et al 1996a) and 0.25
mag for the reddening (Massey et al. 2007). In order to select young
Main Sequence stars with an age of no more than about 100 Myr to
study star complexes, there is a need for colour and magnitude
cutoff limits to define a proper selection slice of the CMD. After
properly fitting theoretical isochrones (Girardi et al. 2002) of
Z=0.004 to the Main Sequence, we adopted the ranges $18\leq V \leq
21$ and $-0.4\leq (B-V)_{0} \leq 0$ for the magnitude and the
colour, respectively. The CMD diagram of NGC 6822, with the selected
slice and its characteristic theoretical isochrones are shown in
Figure \ref{cmd_ellipses}. In the same Figure the photometric
uncertainties are illustrated by the colour and magnitude error bars
of 30 stars randomly selected from the stellar catalogue (see
Section 2).

In order to properly use our data to identify the star complexes, we
produced the stellar density map of the young stars of NGC 6822,
which was constructed by counting stars in bins of
 $23.5^{\prime\prime}\times23.5^{\prime\prime}$ ($57\times57$ pc$^{2}$)
and by applying a $3\times3$
($70.5^{\prime\prime}\times70.5^{\prime\prime}$) mean-smoothing. The
binning mentioned above was chosen after several tests, as the best
compromise between spatial resolution and statistical significance.

To detect the star complexes of NGC 6822 we computed a density
background. By using the pixels of the density map with a value
greater than zero we localize the background calculation so as to be
characteristic of the star formation and to avoid areas outside the
galaxy which artificially lower the background value. The
calculation of the background was carried out by using the
$\sigma-clipping$ method with respect to the median value. The
median and standard deviation values are initially computed for all
the (non zero) pixels. Then, pixels with values higher than
$median+3\sigma$ are excluded and median and $\sigma$ are
recalculated. This procedure continues until convergence, usually
after very few iterations. The final $median+3\sigma$ value is
considered to be the minimum density cutoff limit.

In order to map the star complexes of NGC 6822, we fitted ellipses
to the external isopleth of every distinct structure or substructure
above the minimum density cutoff limit. The isopleths were drawn
from this limit to the maximum value of the density map with a step
of one $\sigma$ and a density range of more than two $\sigma$, to
avoid the selection of random or controversial concentrations. In
some cases optimization led us to the compromise of taking into
account two subsequent isopleths to fit an ellipse. In Figure
\ref{cmd_ellipses} the ellipses that represent the star complexes of
each galaxy are over plotted on the isopleths. The center
coordinates and major axis length of the ellipses are considered to
be the position and size of the star complexes they outline,
respectively. Table \ref{listsfr6822} lists the center coordinates
($\alpha_{2000}$, $\delta_{2000}$) and the size of each one of the
detected star complexes of NGC 6822, sorted in descending order of
the right ascension. They are conventionally named as NGC6822-SC-i,
where SC stands for Star Complex and i ranges from 1 to 24.

\begin{figure*}[h]
   \centering
   \includegraphics[width=.33\textwidth]{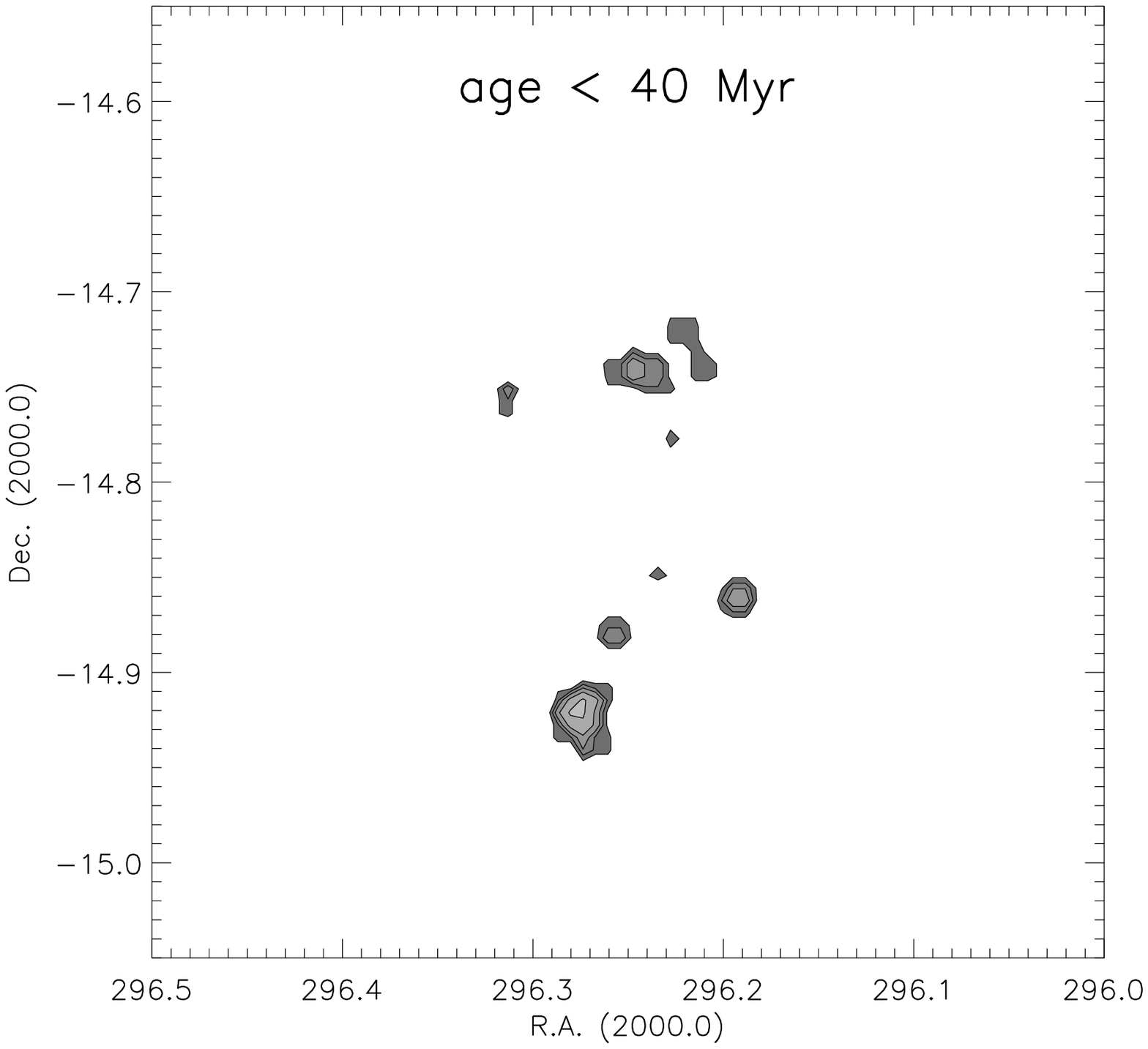}
   \includegraphics[width=.33\textwidth]{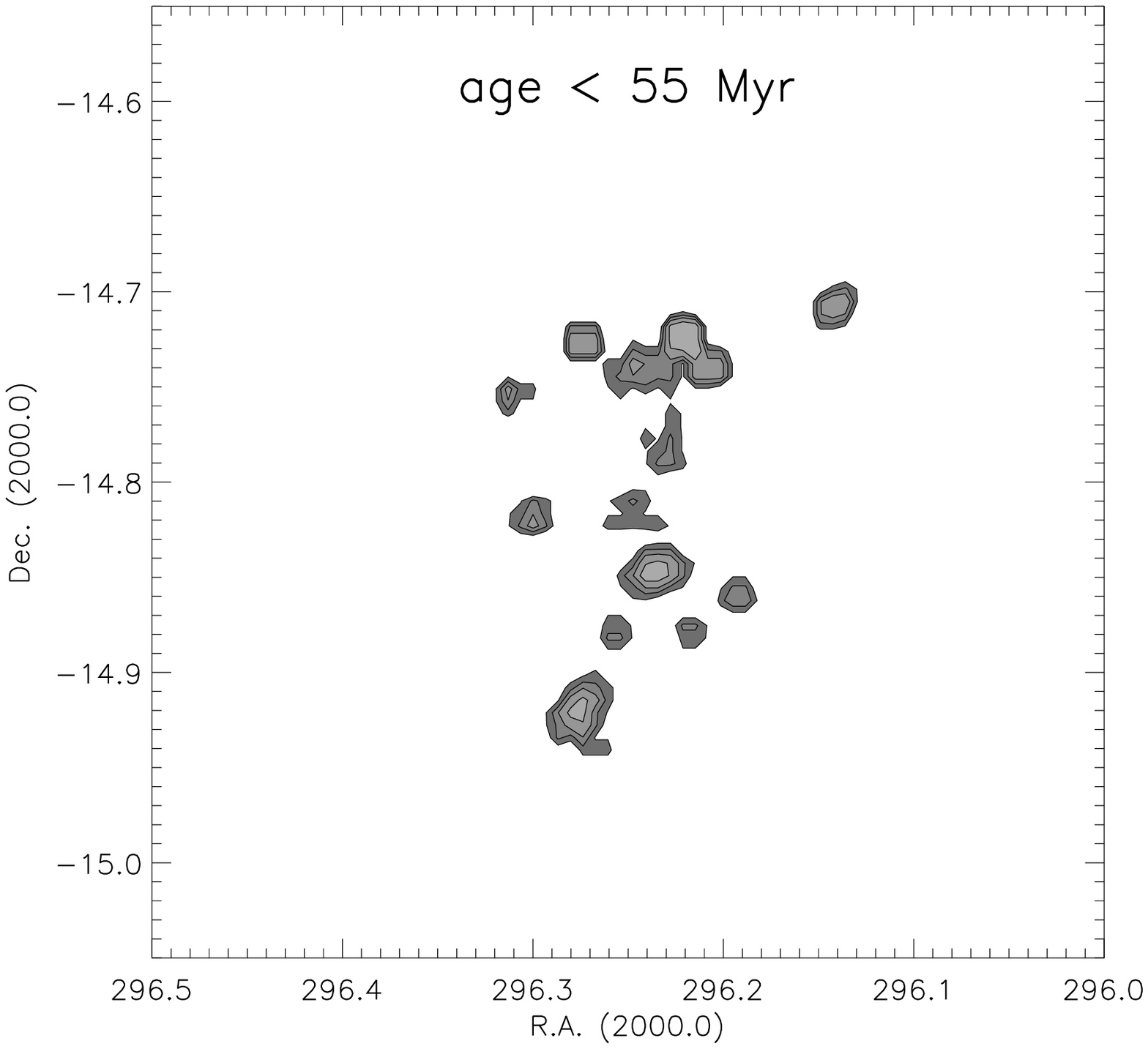}
   \includegraphics[width=.33\textwidth]{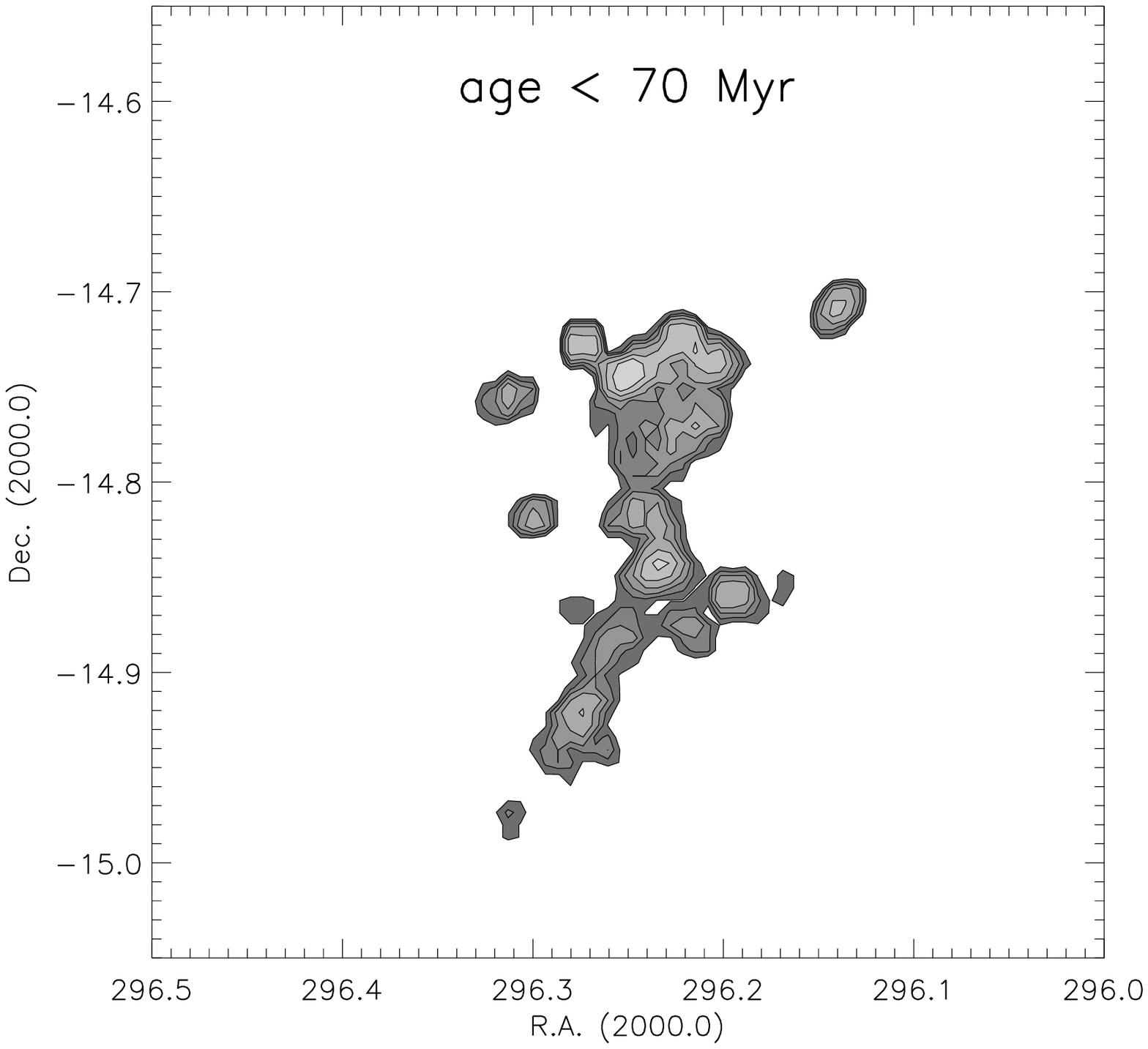}
   \includegraphics[width=.33\textwidth]{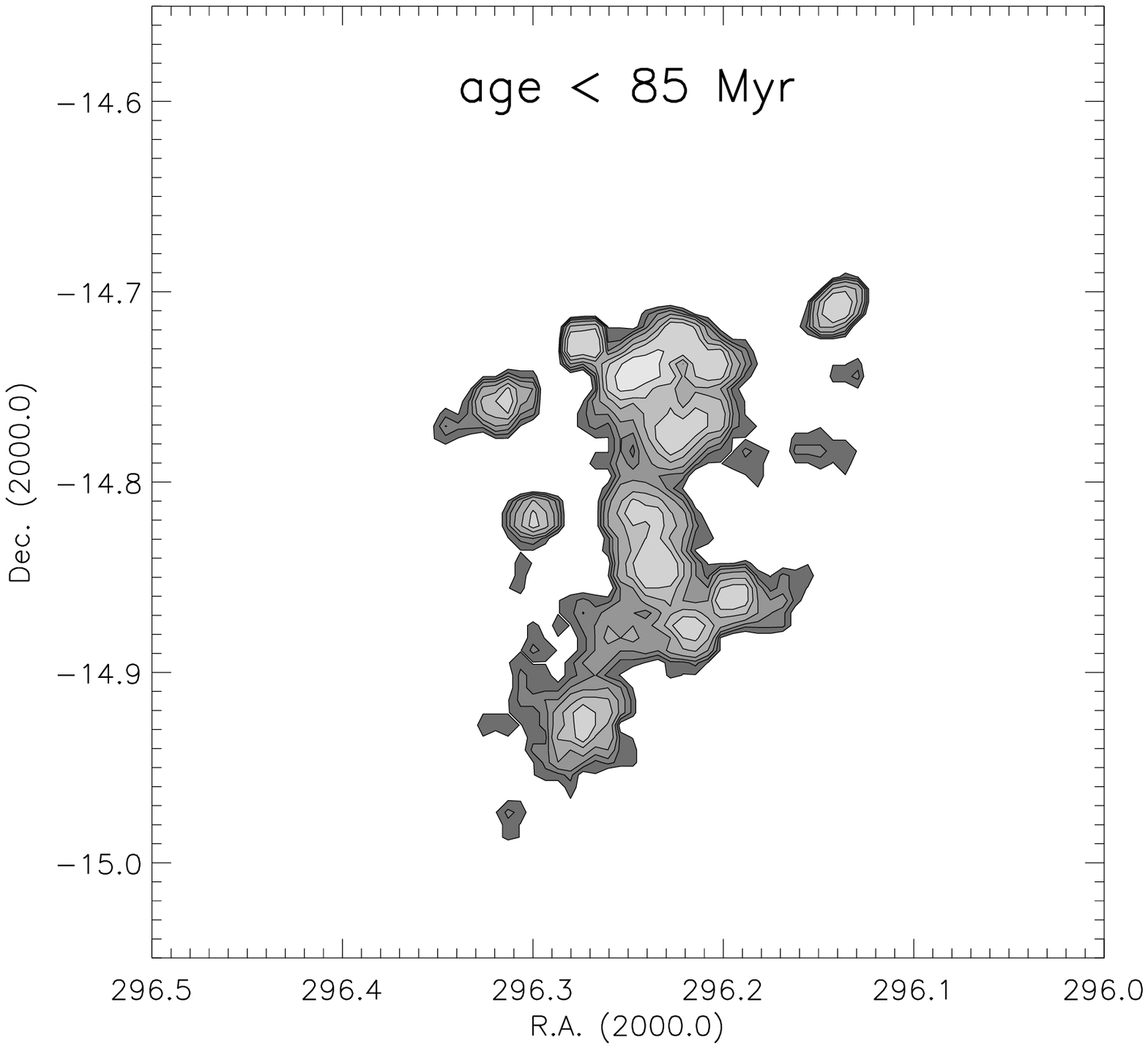}
   \includegraphics[width=.33\textwidth]{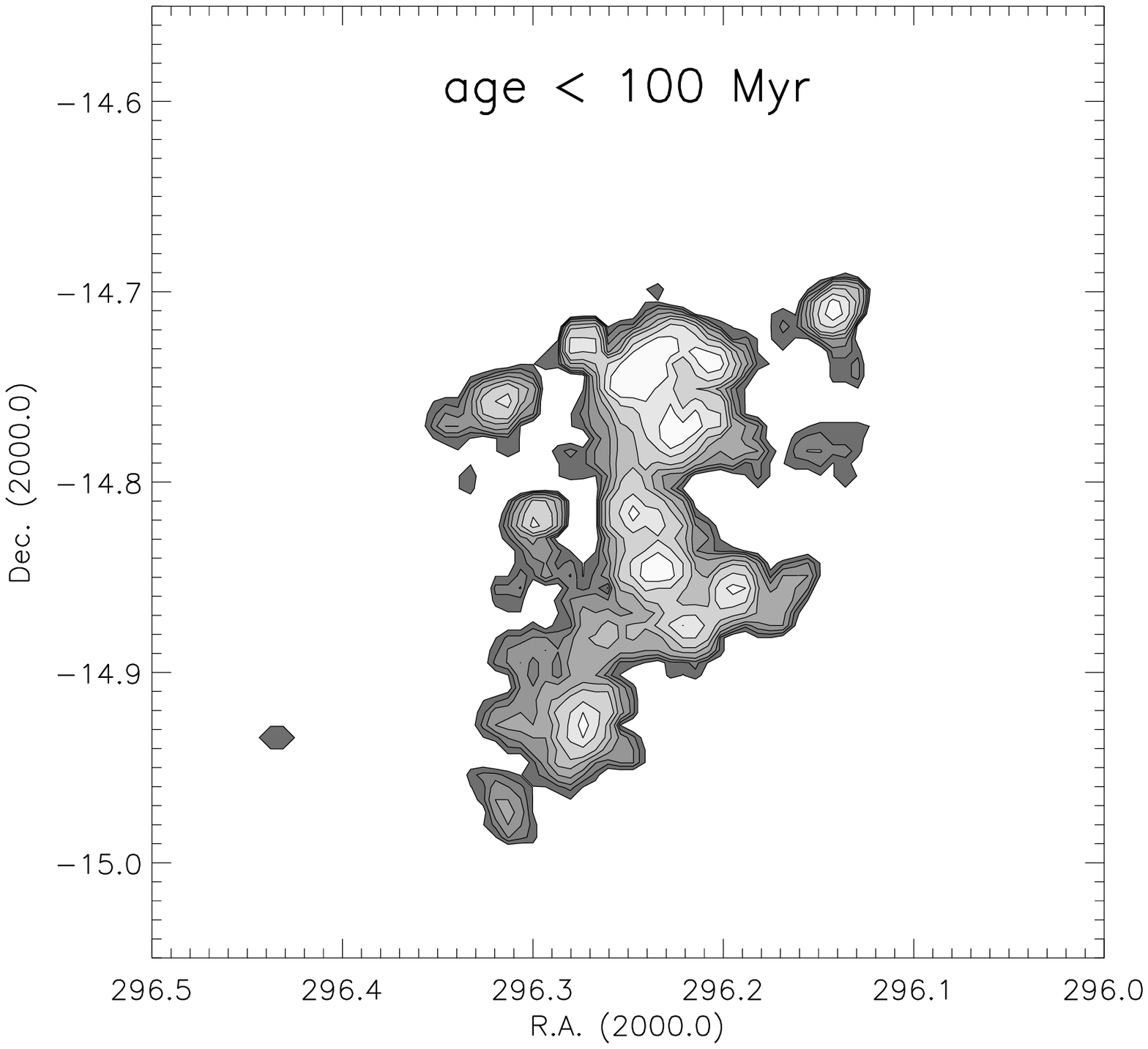}
   \includegraphics[width=.33\textwidth]{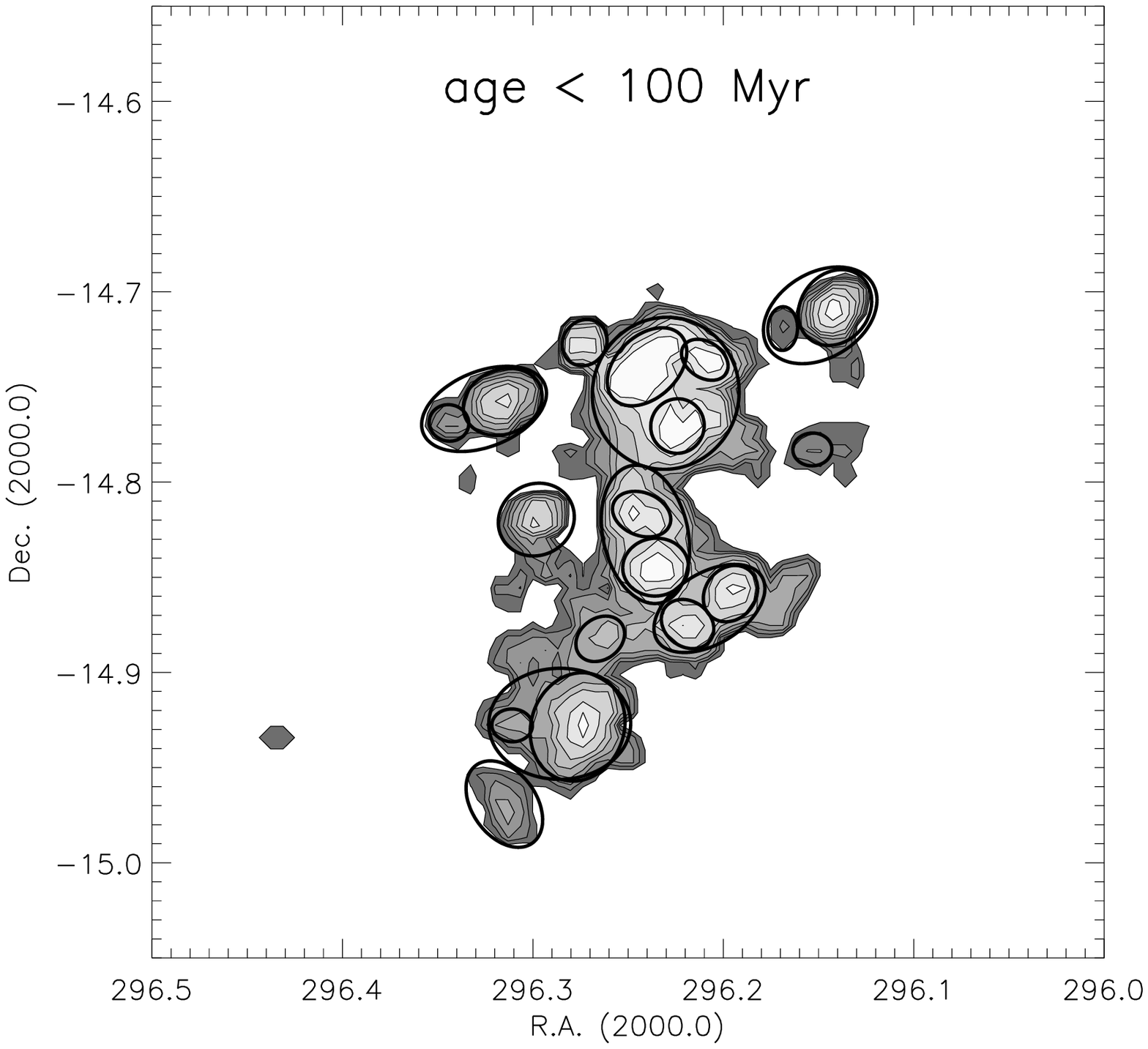}
      \caption{Spatial distribution of young Main Sequence stars of different
      maximum age. The last plot shows the ellipses outlining the detected star complexes of NGC 6822. (Table
        \ref{listsfr6822}).
              }
         \label{cmd_subslices}
   \end{figure*}

Figure \ref{cmd_subslices} shows the spatial distribution of young
Main Sequence stars of different maximum age, ranging from 40 Myr to
100 Myr in time bins of 15 Myr. The last plot shows the ellipses
outlining the detected star complexes of NGC 6822. Structures
younger than $\sim$ 55 Myr are clearly less extended than structures
containing stars with age up to $\sim$ 100 Myr.

\begin{table}
\caption{List of NGC 6822 detected star complexes, in descending
order of the right ascension. Their position and size come from the
center coordinates and major axis length of the ellipses fitted on
them, respectively.} \label{listsfr6822} \centering
\begin{tabular}{l c c c c}
\hline\hline
Name & $\alpha_{2000}$    & $\delta_{2000}$    & Size \\
       & (h m s) & (\degr \ \arcmin \ \arcsec) & (pc)       \\
\hline
NGC6822-SC-1 &   19 45 22.58 &  -14 46 08.15 &  182 \\
NGC6822-SC-2 &   19 45 18.16 &  -14 45 41.62 &  600 \\
NGC6822-SC-3 &   19 45 15.80 &  -14 45 28.33 &  372 \\
NGC6822-SC-4 &   19 45 15.60 &  -14 58 09.05 &  440 \\
NGC6822-SC-5 &   19 45 14.62 &  -14 55 40.15 &  186 \\
NGC6822-SC-6 &   19 45 11.57 &  -14 49 10.96 &  356 \\
NGC6822-SC-7 &   19 45 08.62 &  -14 55 37.20 &  650 \\
NGC6822-SC-8 &   19 45 06.36 &  -14 55 43.11 &  504 \\
NGC6822-SC-9 &   19 45 05.48 &  -14 43 36.30 &  220 \\
NGC6822-SC-10 &   19 45 03.52 &  -14 52 56.50 &  240 \\
NGC6822-SC-11 &   19 44 58.30 &  -14 49 00.62 &  276 \\
NGC6822-SC-12 &   19 44 57.81 &  -14 49 38.96 &  643 \\
NGC6822-SC-13 &   19 44 57.62 &  -14 44 21.98 &  425 \\
NGC6822-SC-14 &   19 44 56.54 &  -14 50 40.89 &  302 \\
NGC6822-SC-15 &   19 44 55.26 &  -14 45 12.13 &  699 \\
NGC6822-SC-16 &   19 44 53.78 &  -14 46 14.05 &  251 \\
NGC6822-SC-17 &   19 44 52.51 &  -14 52 28.49 &  256 \\
NGC6822-SC-18 &   19 44 50.34 &  -14 44 08.74 &  232 \\
NGC6822-SC-19 &   19 44 49.75 &  -14 51 59.00 &  561 \\
NGC6822-SC-20 &   19 44 47.10 &  -14 51 31.00 &  275 \\
NGC6822-SC-21 &   19 44 40.52 &  -14 43 09.73 &  200 \\
NGC6822-SC-22 &   19 44 36.78 &  -14 46 58.26 &  179 \\
NGC6822-SC-23 &   19 44 35.79 &  -14 42 44.68 &  561 \\
NGC6822-SC-24 &   19 44 33.93 &  -14 42 29.95 &  371 \\
\hline
\end{tabular}
\end{table}

\section{Size distribution of star complexes - Comparison with the Magellanic Clouds}

\begin{figure*}
   \centering
   \includegraphics[width=.32\textwidth]{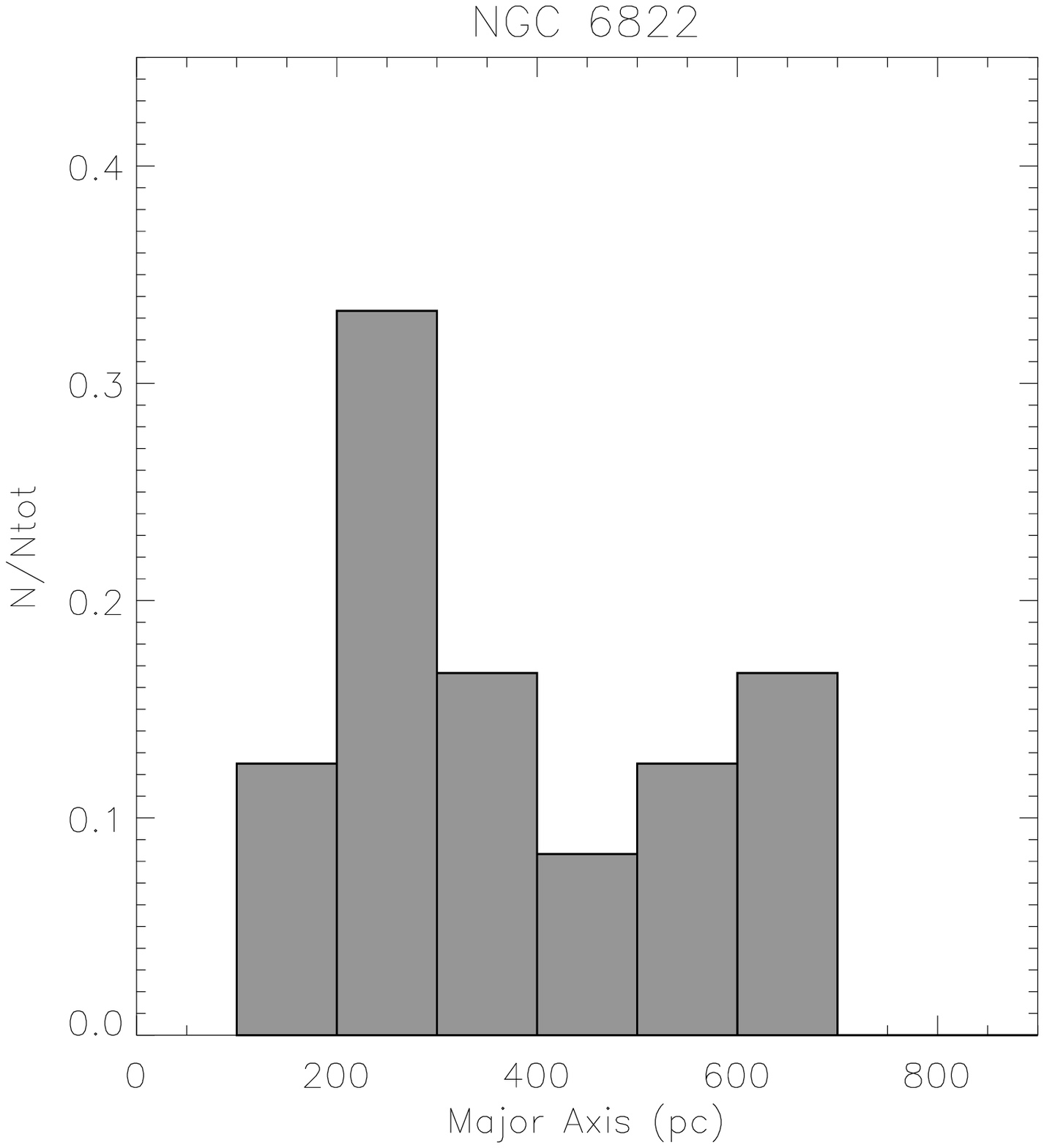}
   \includegraphics[width=.32\textwidth]{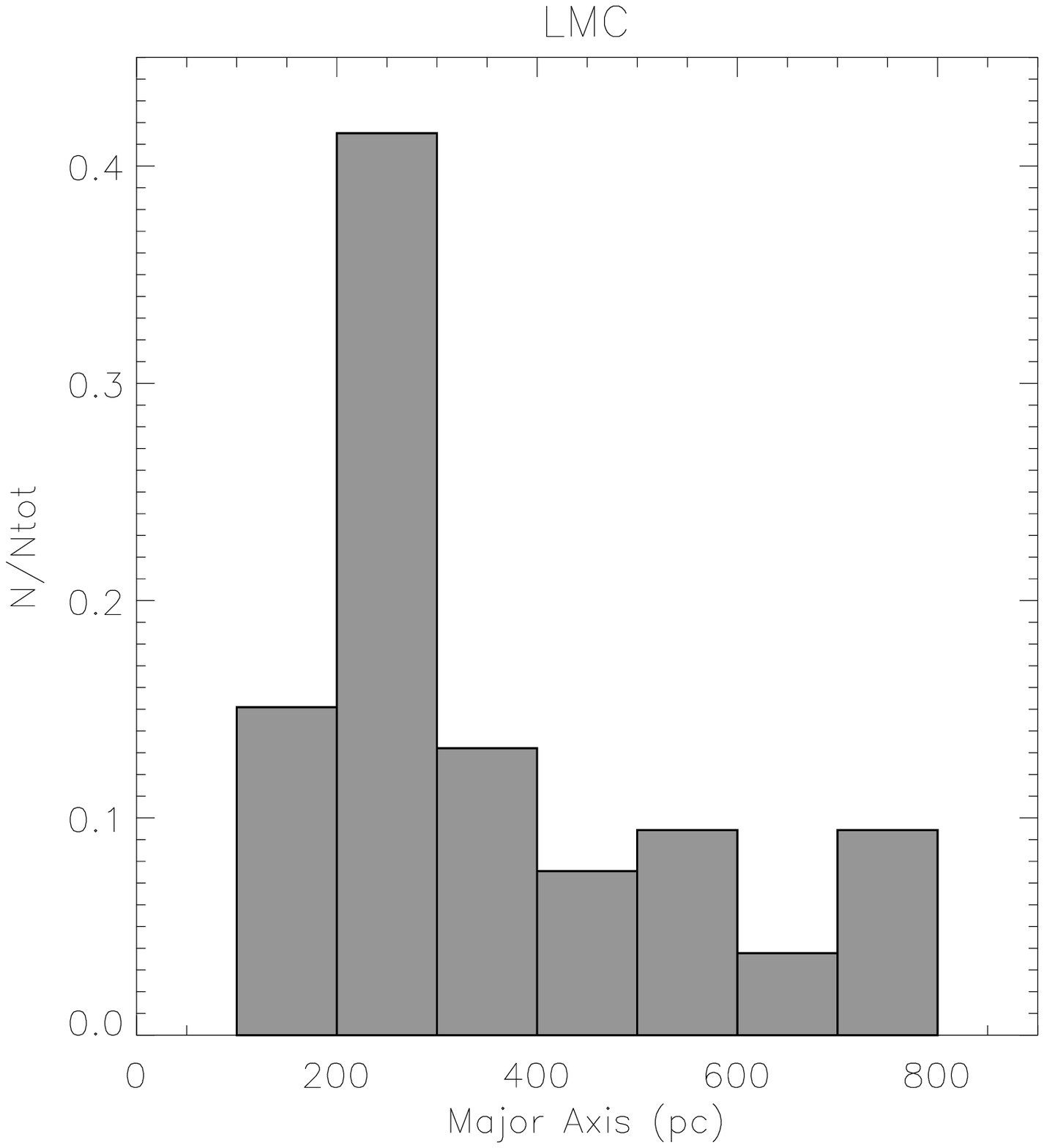}
   \includegraphics[width=.32\textwidth]{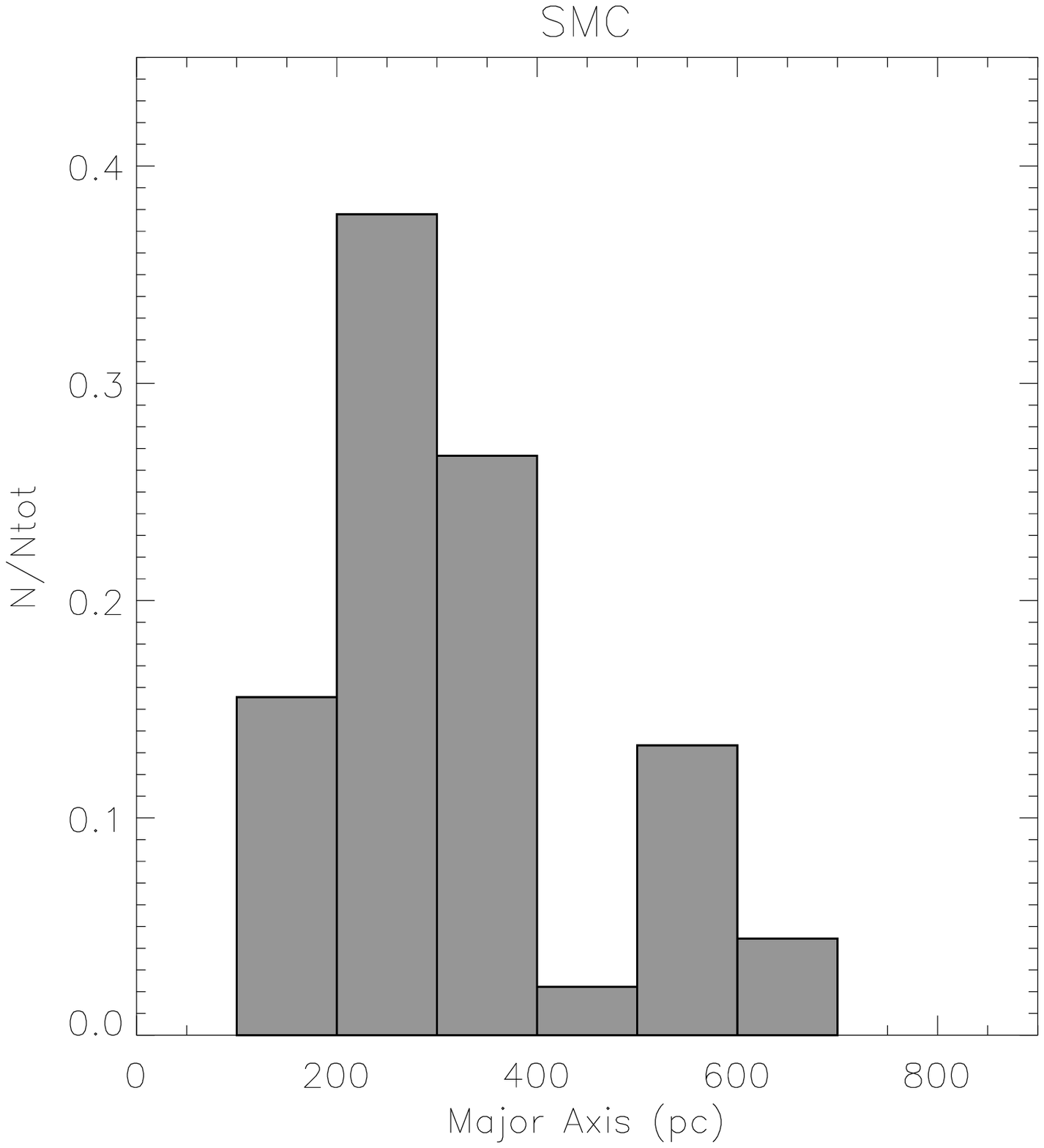}
      \caption{Histograms of the size distribution of detected star forming regions
      of NGC 6822 and the Magellanic Clouds, with a bin size of 100 pc.
              }
         \label{allregions}
   \end{figure*}

\begin{figure*}
   \centering
   \includegraphics[width=.32\textwidth]{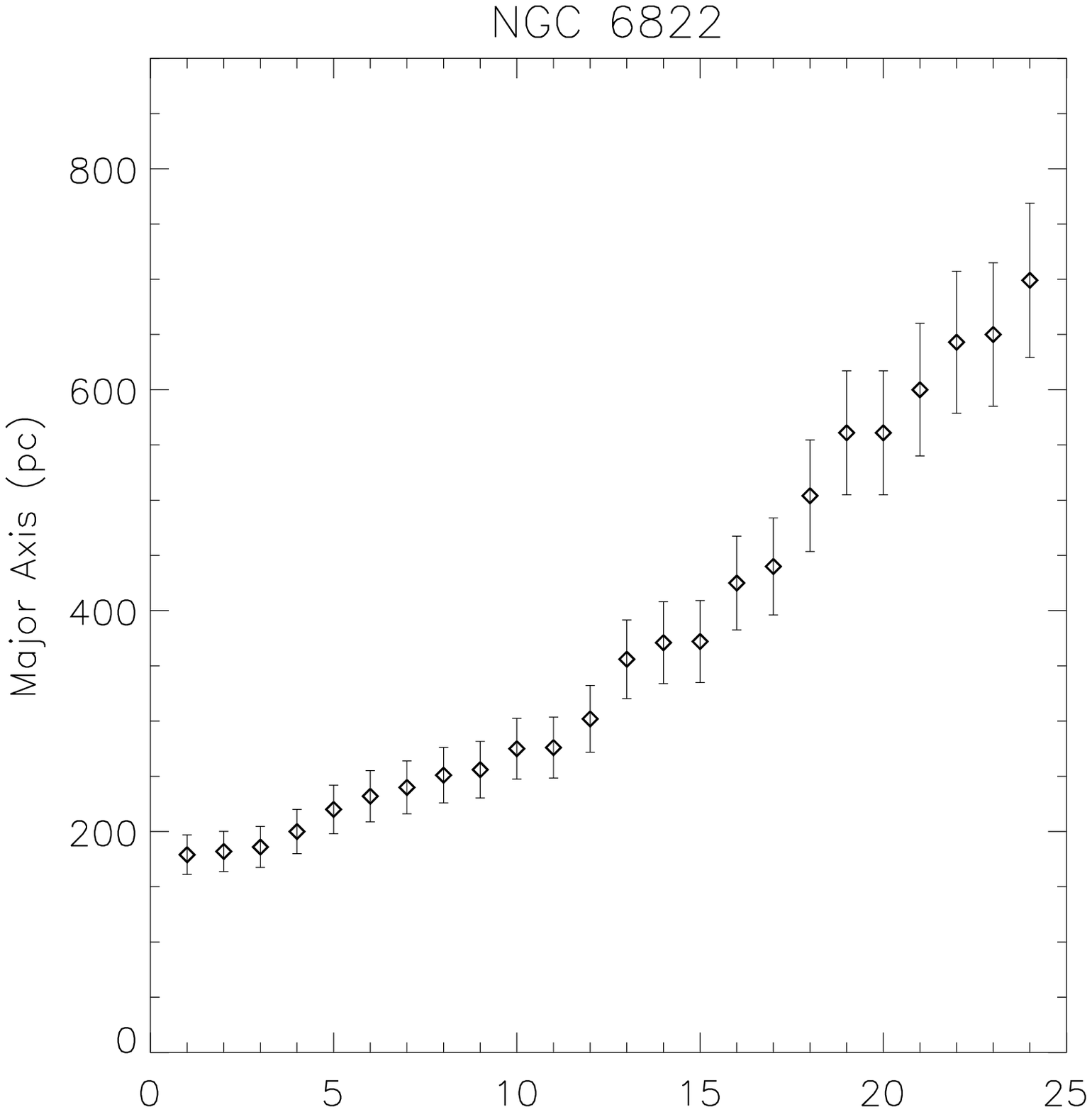}
   \includegraphics[width=.32\textwidth]{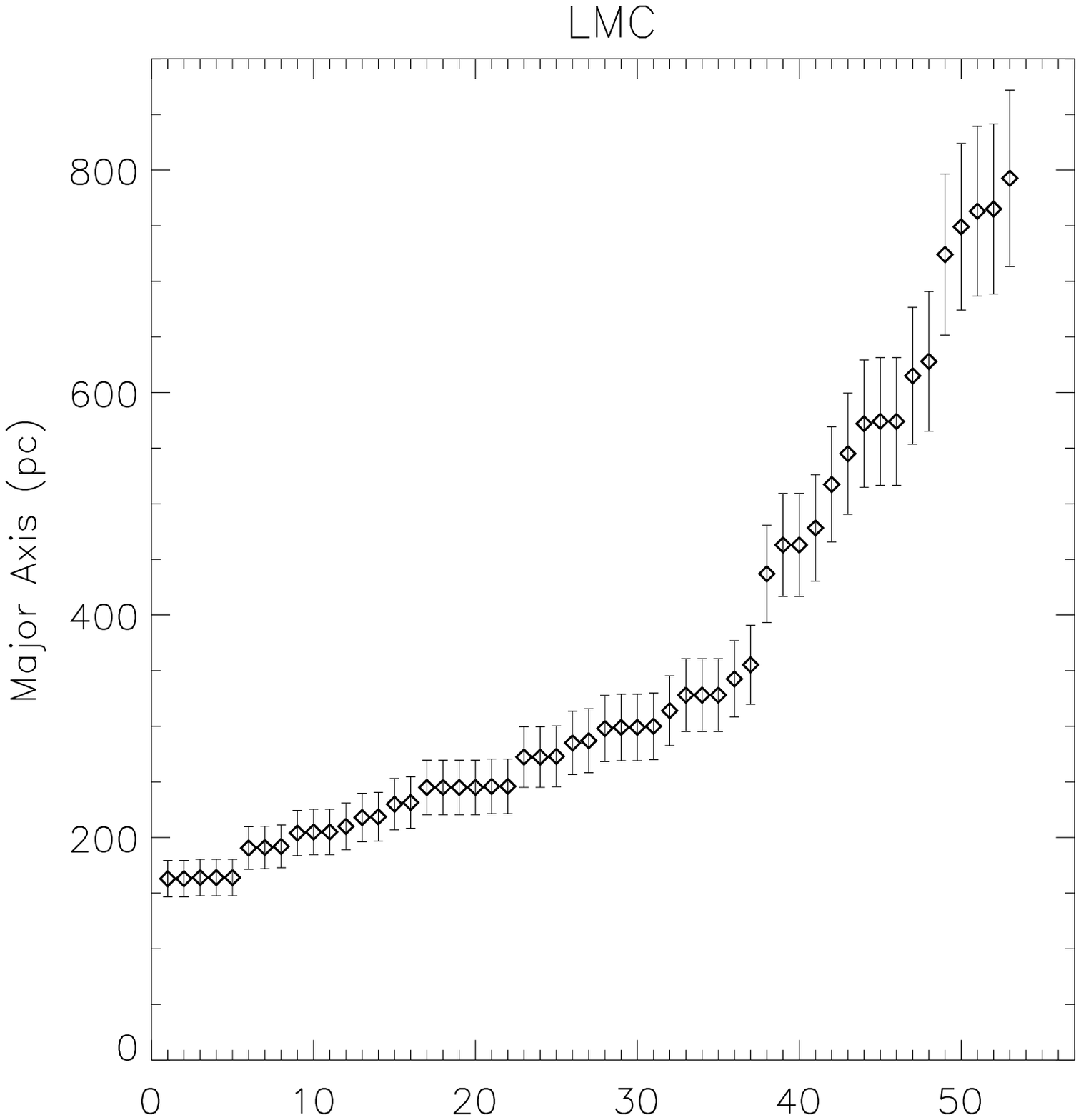}
   \includegraphics[width=.32\textwidth]{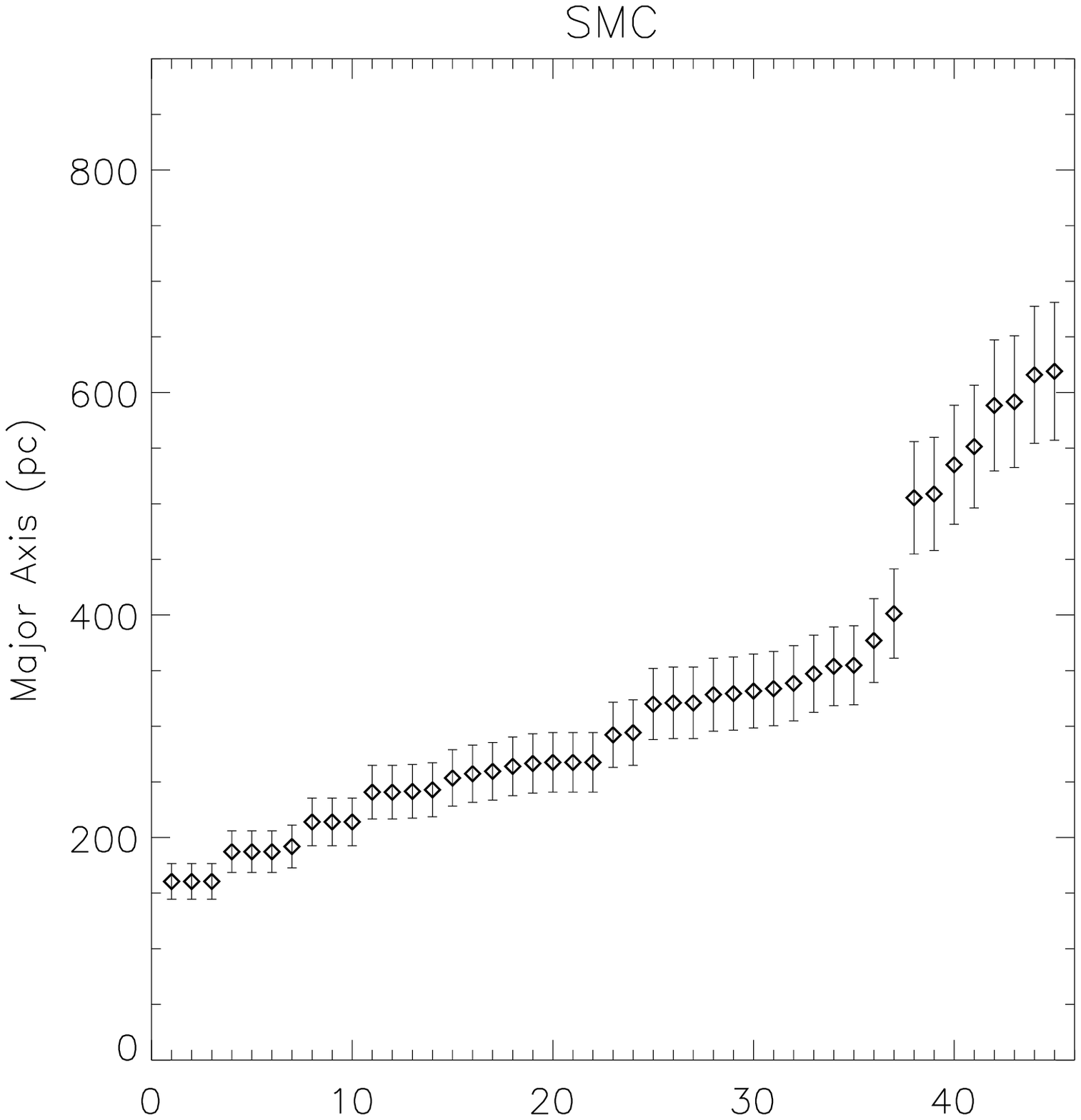}
      \caption{Sizes of the detected star forming regions
      of NGC 6822 and the Magellanic Clouds, sorted in ascending order.
              }
         \label{allpoints}
   \end{figure*}

Since the star complexes have been detected, what is examined next
is the distribution of their size (as it was defined in Section 3).
In Figure \ref{allregions} the histogram of the size distribution of
detected star complexes of NGC 6822 is shown, with the same plots
for the Large Magellanic Cloud (Maragoudaki et al. 1998; Livanou et
al. 2006), based on optical observations, and the Small Magellanic
Cloud (Livanou et al. 2007), based on infrared observations, with a
common bin size of 100 pc. A few regions of the MCs with size $>$
900 pc have been excluded for reasons of homogeneity and statistical
significance. The MCs were chosen for comparison for two reasons:
\emph{(a)} They are our nearest neighbors in the Local Group and
their star complexes are well investigated, and \emph{(b)} these
complexes were detected, in principal, with the same method we use
in this paper. In Figure \ref{allpoints} the sizes of the detected
star complexes of NGC 6822 and the MCs are presented, sorted in
ascending order. The error for each size value was assumed to be the
10\% of this value for NGC 6822 (this work) and the Magellanic
Clouds (Livanou et al 2006, 2007). These plots aim to contribute in
a supplementary way to the histograms described above, since they
could better reveal the limits, if any, between different size
groups. When carefully observing these diagrams, two main size
groupings can be distinguished for the three galaxies. The first
group ranges from $\sim$ 150 pc to 300-400 pc and the second up to
$\sim$ 800 pc. We briefly refer to them as Group I and Group II,
respectively. In order to investigate whether these size groupings
are indicative of preferable size ranges of star formation in these
galaxies, we compared them statistically.

If an apparently distinct group is indeed distinct, it is assumed
that the size values it contains should be distributed in a uniform
random way. Similarly thinking, if we combine two true distinct
groups, which are self-characterized by a uniform random
distribution ($U_{R}$), their combination should not be
characterized by this property any more. If it could maintain this
property, then these two groupings would actually be one, in
contradiction with the hypothesis that they are two different
groups. So, we compared the two main apparent groupings and their
combination with an equal number of uniformly random size values,
limited to the same range, for NGC 6822 and the Magellanic Clouds.
The comparison was repeated a thousand times to limit the impact of
some possible extreme cases of random distributions.

Having to compare two equally unknown data sets of continuous data
as a function of a single variable (the size), the two-tailed
Kolmogorov-Smirnov (K-S) test is the most appropriate. The Null
Hypothesis is that the two data sets are drawn from the same
distribution. We performed these tests and compared the mean value
of the significance level for each pair. Low values of this
probability indicate that the two data sets are significantly
different, while high values indicate that they are probably
consistent with a single distribution. The results for NGC 6822 and
the Magellanic Clouds are listed in Table \ref{ks}.

As an extra control and verification of the procedure described
above we did the following: \emph{(a)} We compared two uniform
random distributions a thousand times to check the mean value of the
significance level, which turned up to be $52.7\%$. This can be
considered as a satisfactory value for two data sets, one being
random, to be drawn from the same distribution. \emph{(b)} We cut
the first grouping of the LMC in half and found no consistency
between these sub groupings ($0\%$ significance level). This is a
sufficient indicator of restriction to the apparent size groupings.

\begin{table}
\caption{Significance levels (\%) of the Kolmogorov-Smirnov test for
NGC 6822 and the Magellanic Clouds. Col.1: Galaxy. Cols 2-4:
Significance levels for the different pairs of size ranges. }
\label{ks} \centering
\begin{tabular}{l c c c c}
\hline\hline
 & Group I & Group II & Group I+II \\
 & $U_{R}$ & $U_{R}$  & $U_{R}$    \\
\hline
NGC 6822 & 72.7 & 72.9 & 26.2 \\
LMC      & 49.7 & 64.3 &  0.2 \\
SMC      & 48.5 & 72.5 &  2.1 \\
\hline
\end{tabular}
\end{table}

The derived values of significance levels show that the apparent
size groupings of the detected star forming regions could be real
distinct groupings: When we compare the size values of Group I with
an equal number of uniformly random size values, limited to the same
range, they seem to be drawn from the same distribution. Similarly,
when we compare the size values of Group II with the corresponding
uniform random size values, again the Null Hypothesis seems to be
correct. This is not the case when we consider Group I and Group II
as one group and compare them with the corresponding uniform random
distribution. Note that Group I is considered to range up to $\sim$
300 - 350 pc for NGC 6822 and up to $\sim$ 350 - 400 pc for the MCs.
Trials of the Kolmogorov-Smirnov test closely around these limits do
not change the results significantly. This difference in the upper
limit of the first Group between NGC 6822 and the MCs could be due
to selection effects (optical or infrared data) and size
determination uncertainties.

\section{Stellar populations}

\begin{figure*}
\centering
\includegraphics[width=0.5\textwidth]{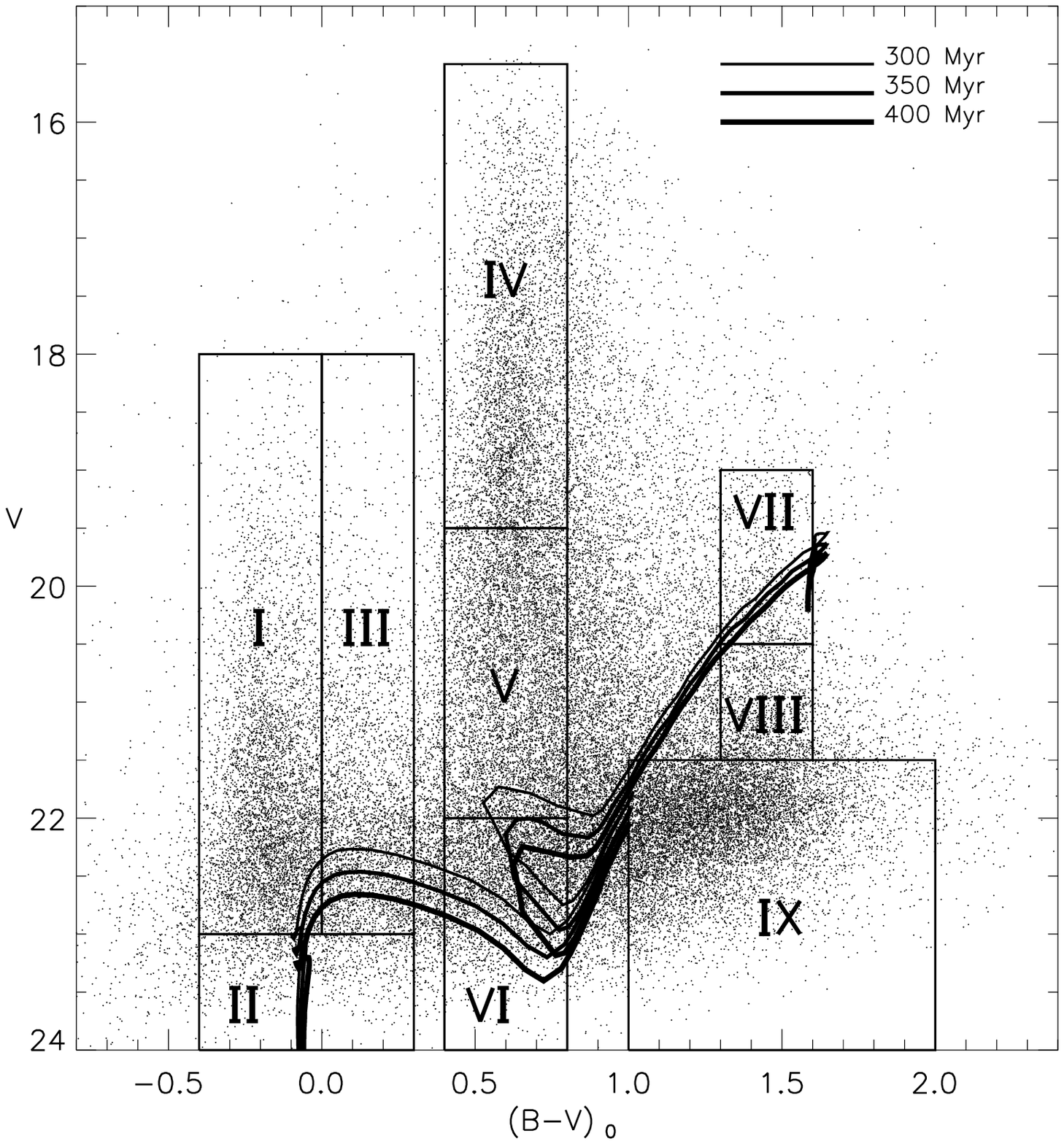}
\caption{The regions of the Colour-Magnitude Diagram of NGC 6822
selected to study the respective stellar populations, and the 300,
350 and 400 Myr theoretical isochrones that roughly separate the
young and the older populations of NGC 6822.} \label{cmdslices}
\end{figure*}

\begin{figure*}
\centering
\includegraphics[width=0.32\textwidth]{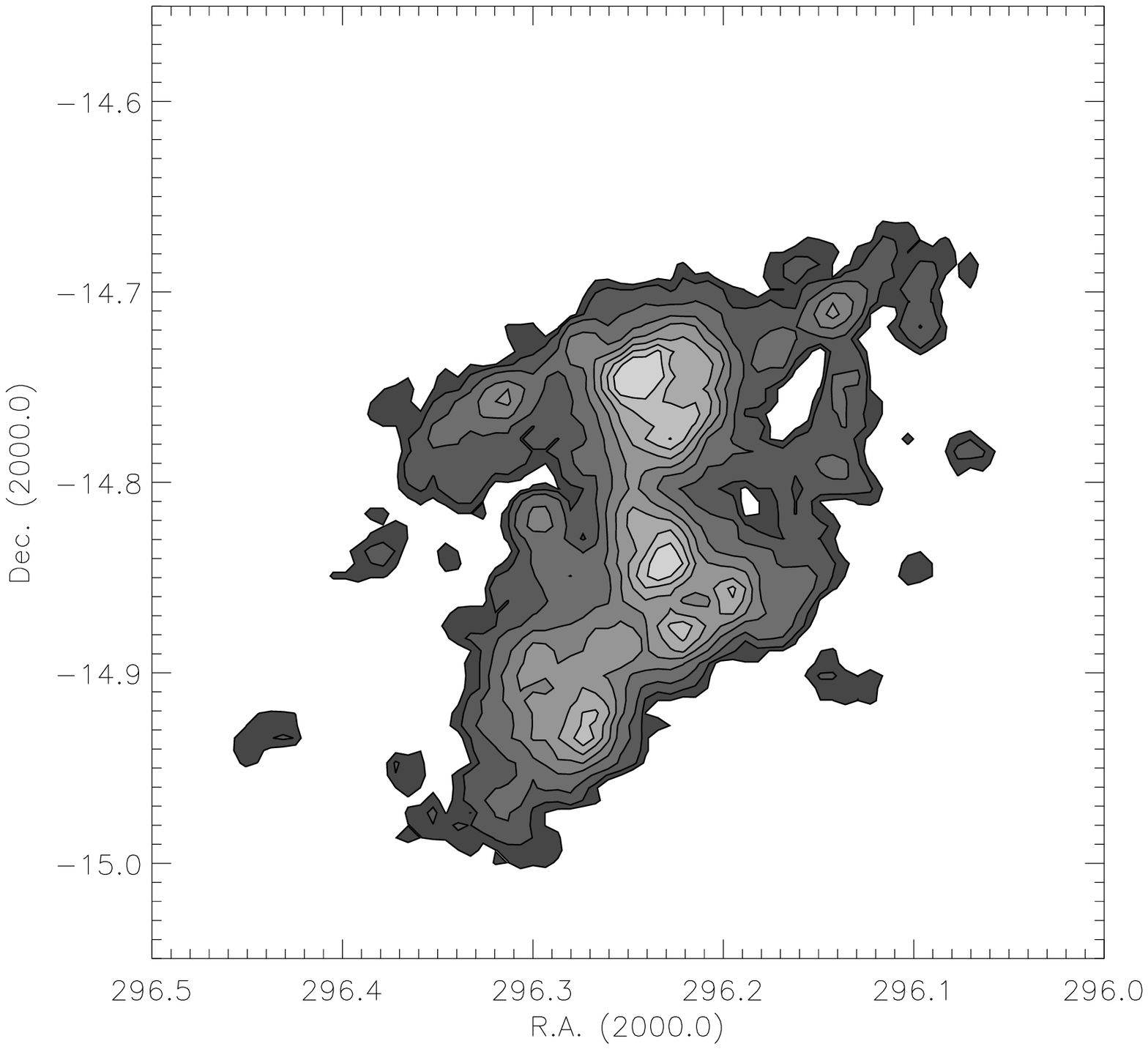}
\includegraphics[width=0.32\textwidth]{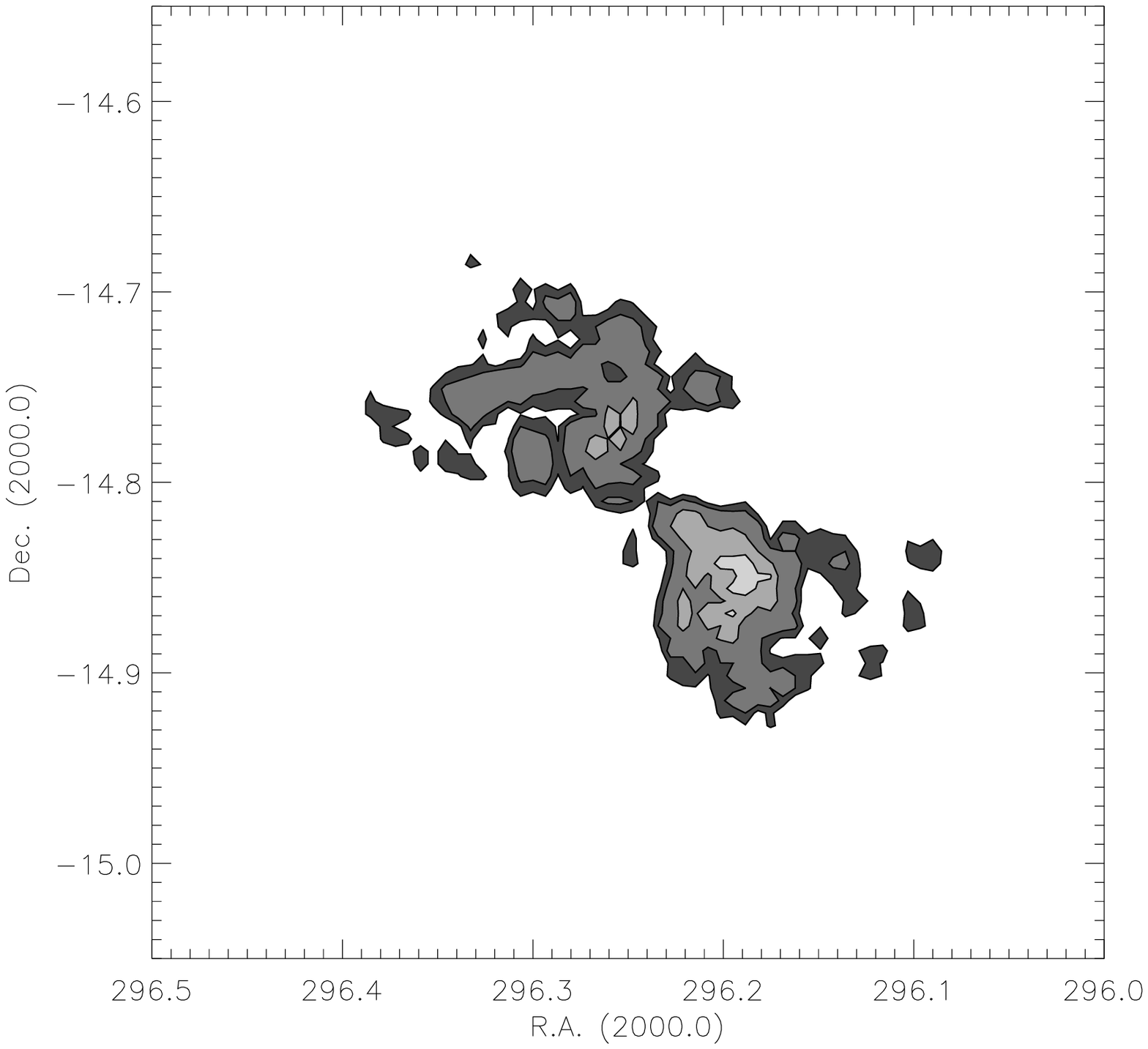}
\includegraphics[width=0.32\textwidth]{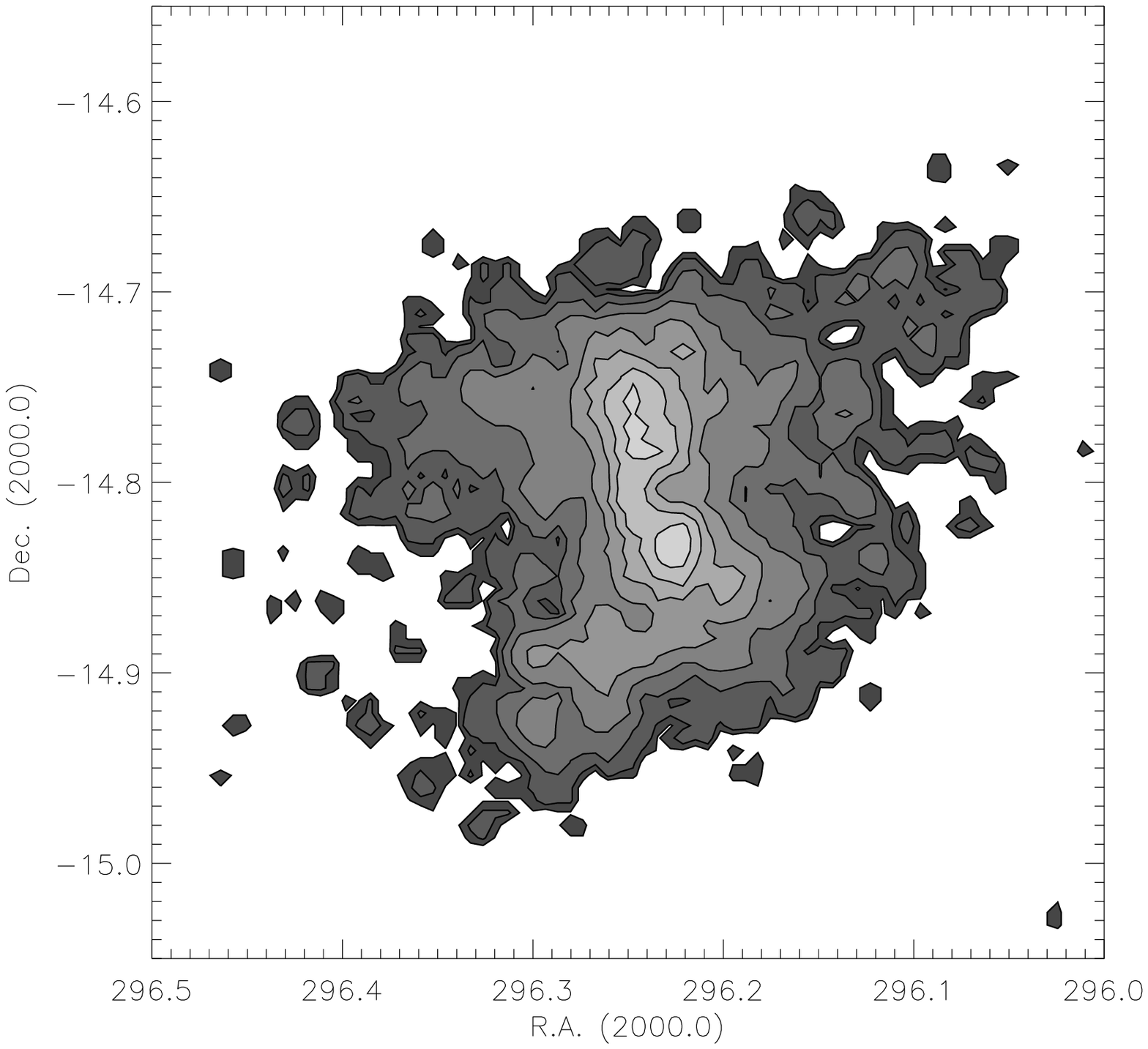}
\includegraphics[width=0.32\textwidth]{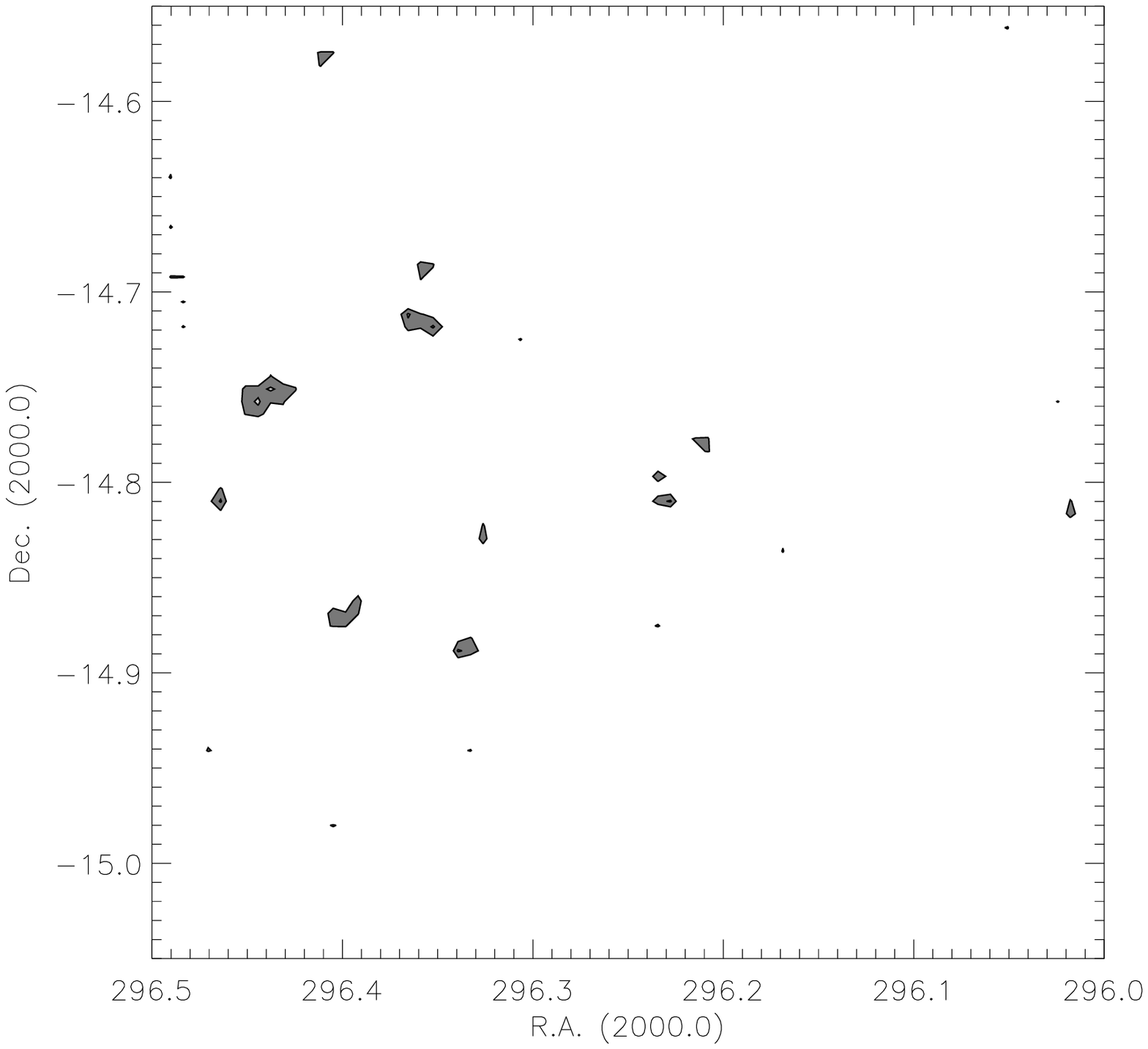}
\includegraphics[width=0.32\textwidth]{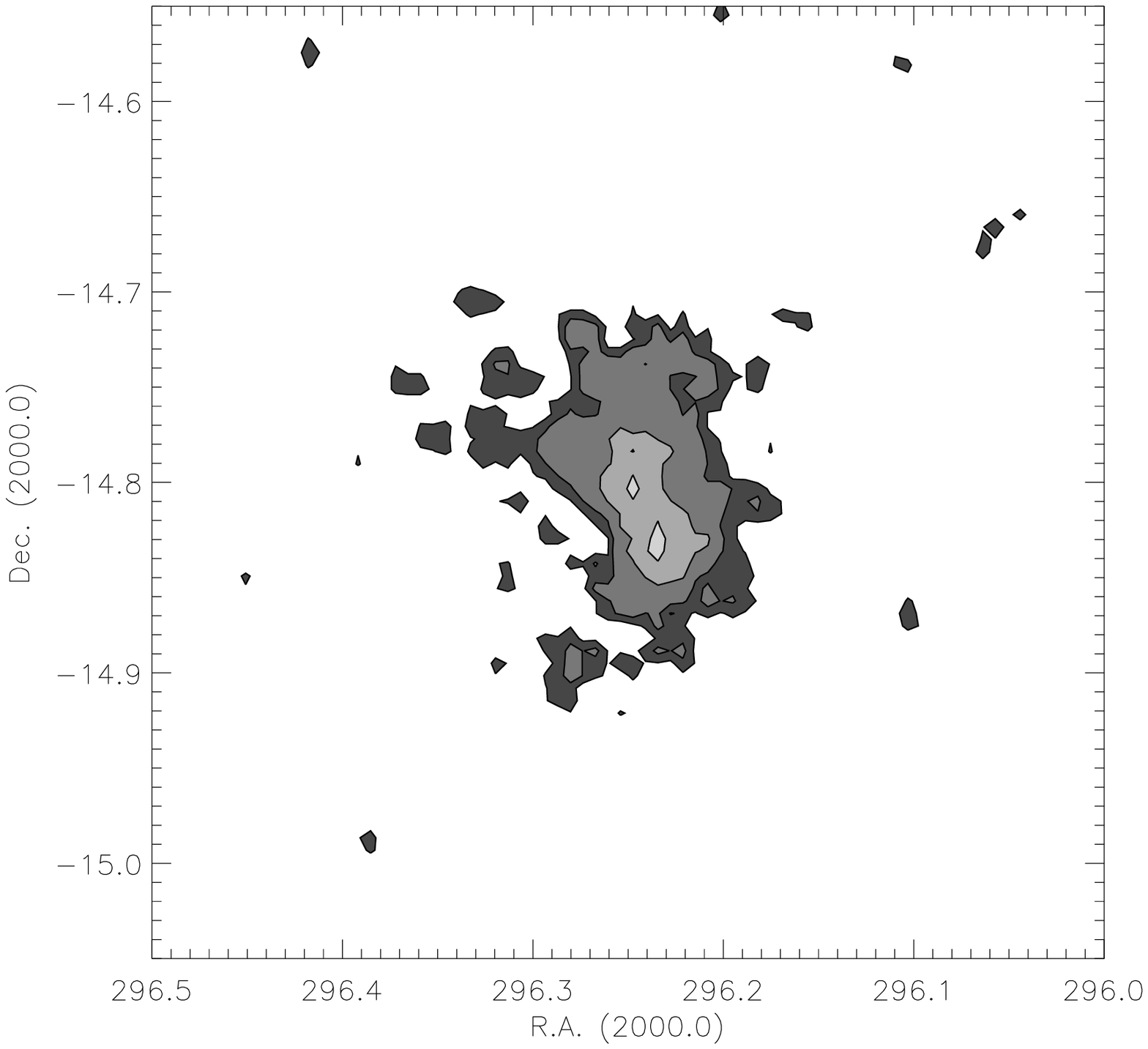}
\includegraphics[width=0.32\textwidth]{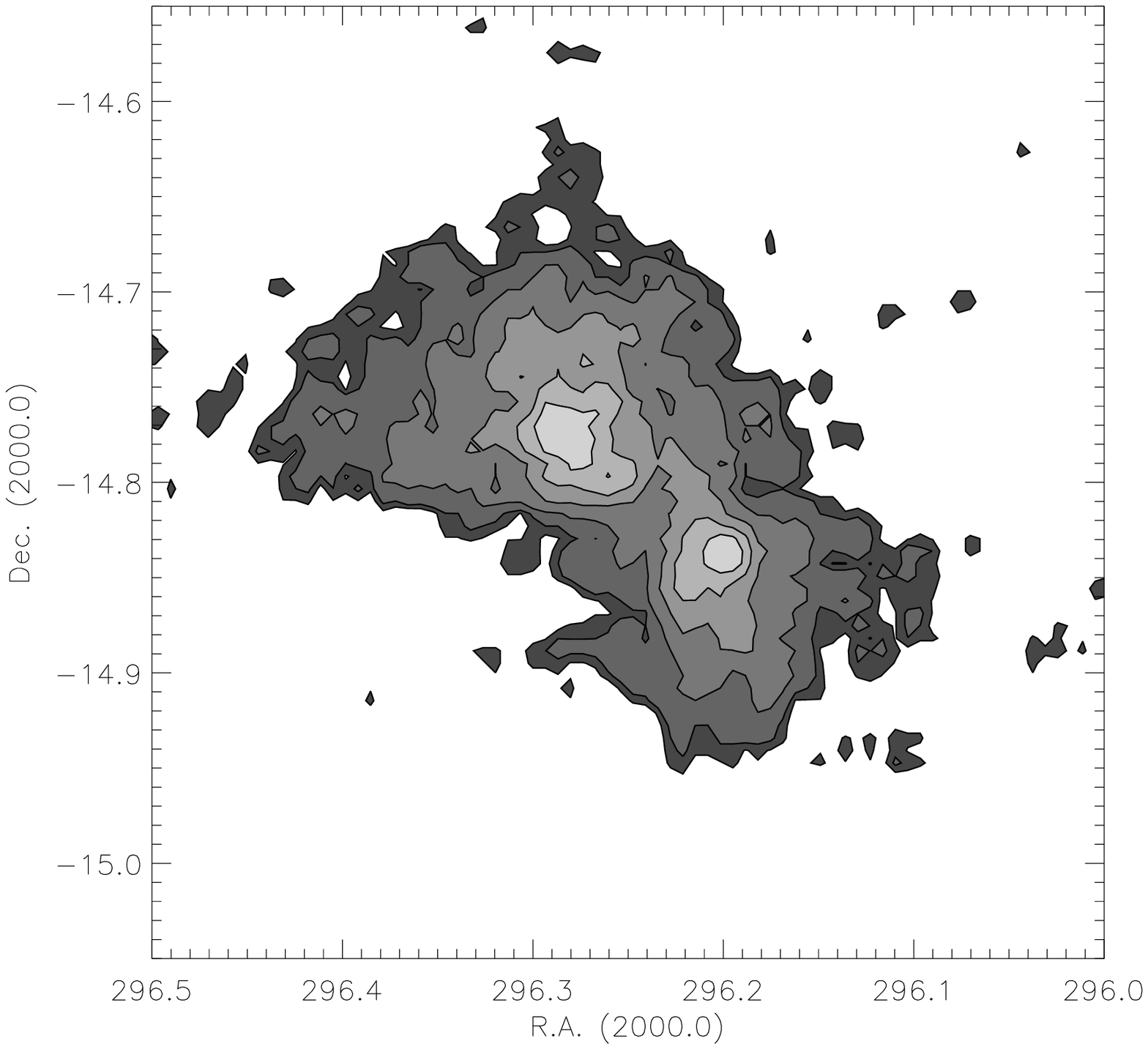}
\includegraphics[width=0.32\textwidth]{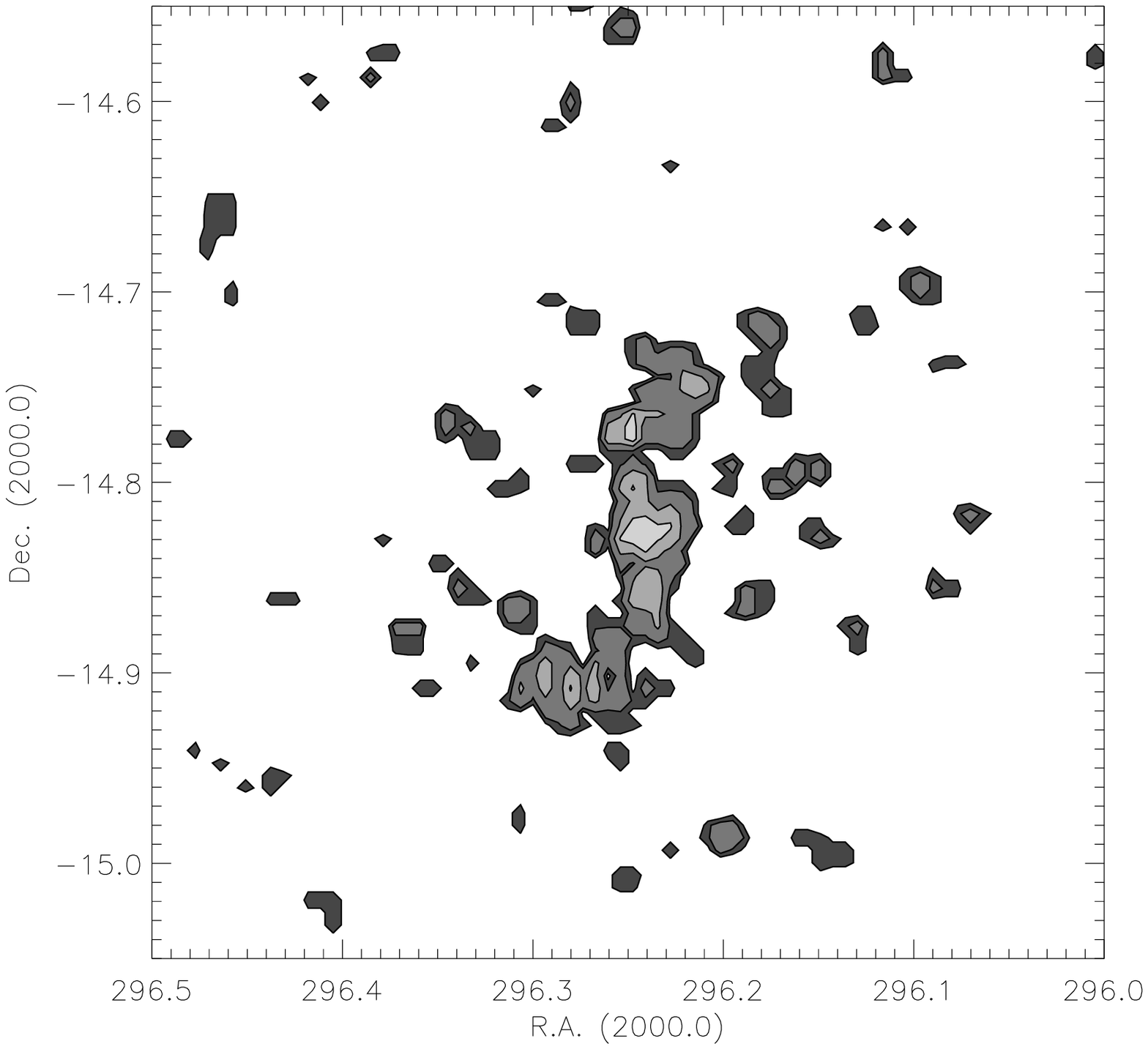}
\includegraphics[width=0.32\textwidth]{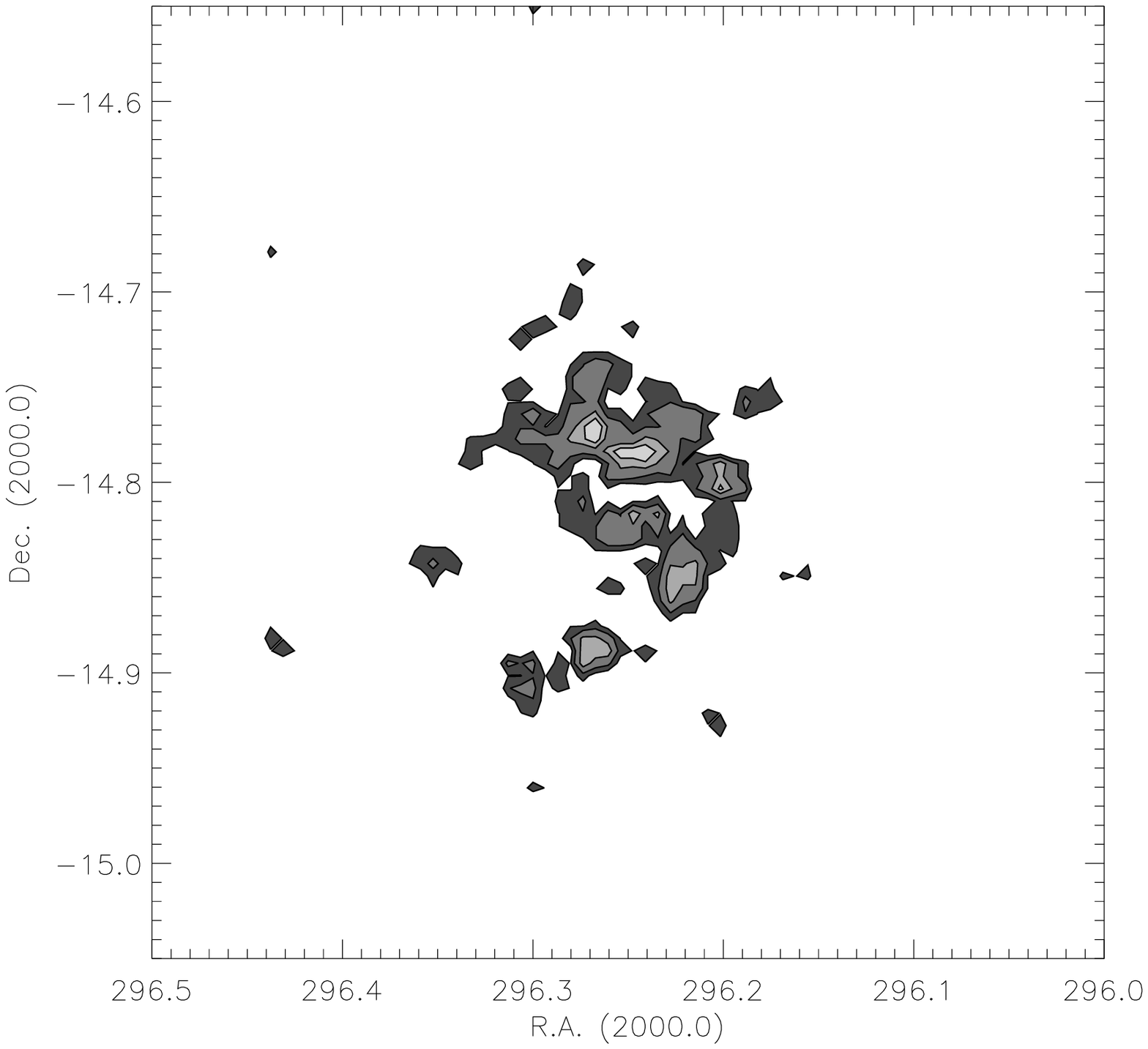}
\includegraphics[width=0.32\textwidth]{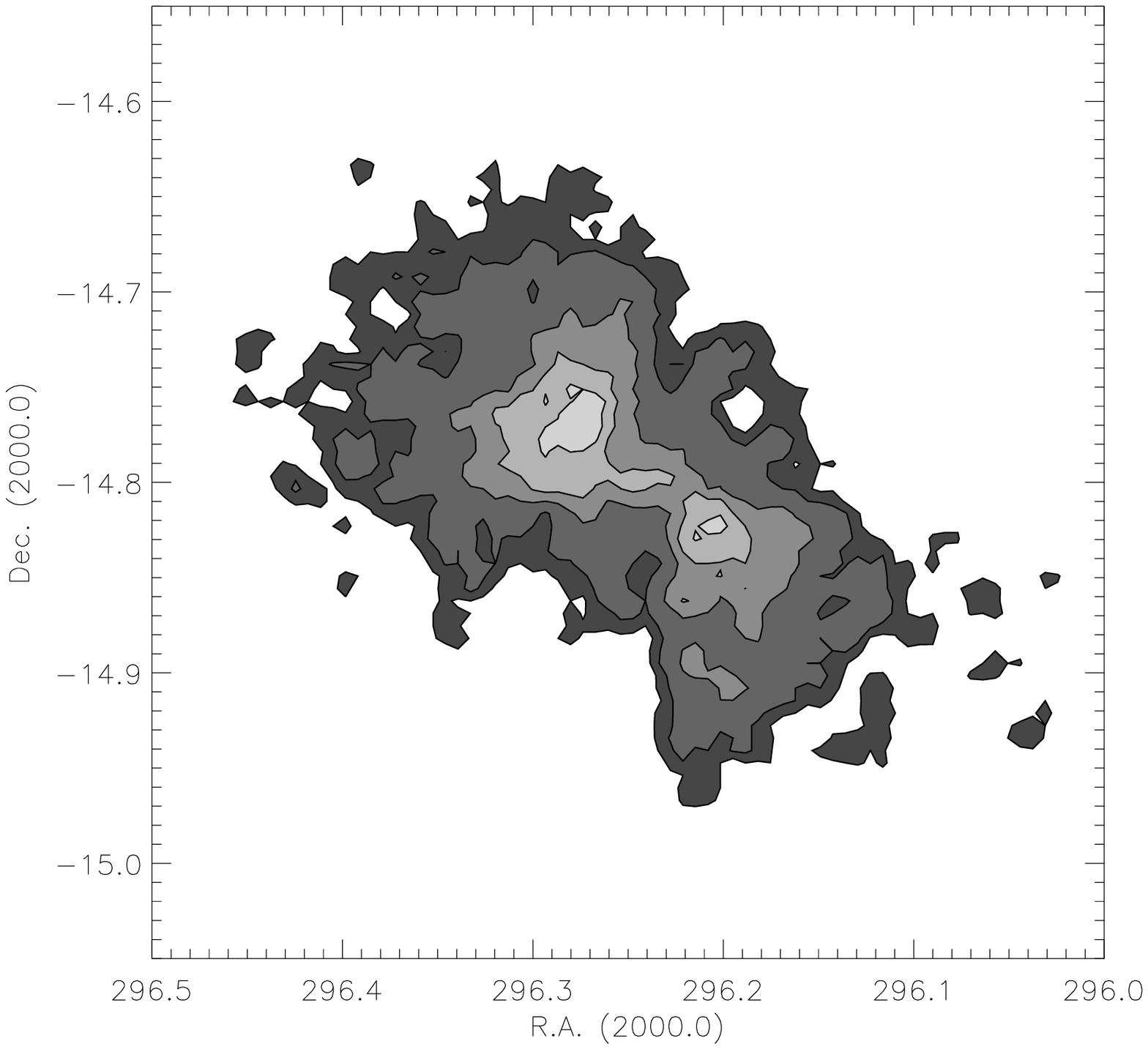}
\caption{Spatial distribution of different stellar populations of
NGC 6822. \textbf{Upper:} From left to right, CMD regions I, II and
III. \textbf{Middle:} From left to right, CMD regions IV, V and VI.
\textbf{Lower:} From left to right, CMD regions VII, VIII and IX.}
\label{9regions}
\end{figure*}

As described in section 3, a proper selection of the young (age
$\leq 100 Myr$) MS stars from the CMD allowed us to produce the
spatial distribution of the star complexes and determine them to
further investigate some of their geometrical properties. Proceeding
the same way, we selected stellar populations of various ages to
compare their spatial distribution. The final selection was the
result of several tests aiming to trace differences in the spatial
distribution of stellar populations, and to cover the most important
features of the Colour-Magnitude Diagram, like the Main Sequence and
the Red Giant Branch. Figure \ref{cmdslices} shows the selected
regions from the CMD outlined on it. The overplotted 300, 350 and
400 Myr theoretical isochrones are used to illustrate the separation
between the young and the older populations of NGC 6822 (see also
Figure \ref{9regions}). Table \ref{regions} lists the colour,
magnitude and age limits and the dominant stellar component of the
selected CMD slices.

\begin{table}
\caption{Colour, magnitude and age limits and dominant stellar
component of the selected Colour-Magnitude Diagram regions.}
\label{regions} \centering
\begin{tabular}{ccccc} \hline \hline

Selected & $V$ & $(B-V)_{\circ}$ & age & Stellar\\
Region & (mag) & (mag) & (Myr) & Content\\
\hline
  I    & $18.0$ - $23.0$   & $-0.4$ - $0.0$  & $\leq 320$  & MS       $^{\mathrm{a}}$ \\
  II   & $23.0$ - $24.0$   & $-0.4$ - $0.3$  & $\leq 630$  & MS       $^{\mathrm{a}}$ \\
  III  & $18.0$ - $23.0$   & $0.0$ - $0.3$   & $60-400$    & Post MS  $^{\mathrm{b}}$ \\
  IV   & $15.5$ - $19.5$   & $0.4$ - $0.8$   & $10-80$     & MW stars $^{\mathrm{c}}$ \\
  V    & $19.5$ - $22.0$   & $0.4$ - $0.8$   & $80-320$    & Post MS  $^{\mathrm{b}}$ \\
  VI   & $22.0$ - $24.0$   & $0.4$ - $0.8$   & $320-500$   & Post MS  $^{\mathrm{b}}$ \\
  VII  & $19.0$ - $20.5$   & $1.3$ - $1.6$   & $130-630$   & MW stars $^{\mathrm{c}}$ \\
  VIII & $20.5$ - $21.5$   & $1.3$ - $1.6$   & $630-3160$  & MW stars $^{\mathrm{c}}$ \\
  IX   & $21.5$ - $24.0$   & $1.0$ - $2.0$   & $\geq 790$  & RGs      $^{\mathrm{d}}$ \\\hline
\end{tabular}
\begin{list}{}{}
\item[$^{\mathrm{a}}$] Main Sequence stars.
\item[$^{\mathrm{b}}$] Post Main Sequence stars.
\item[$^{\mathrm{c}}$] Mostly Milky Way stars. The age range in Col. 4
is referring to NGC 6822 stars.
\item[$^{\mathrm{d}}$] Red Giants.
\end{list}
\end{table}

Following the steps described in Section 3, density maps were
constructed by counting stars in bins of
 $23.5^{\prime\prime}\times23.5^{\prime\prime}$ and by applying a
 $3\times3$
mean-smoothing. Isopleths were drawn for densities higher than the
background level, computed using the $\sigma-clipping$ method in
respect with the median value. Figure \ref{9regions} quotes the
spatial distribution of the different stellar populations.

The slices I, III and V, containing young stars with an age of less
than roughly 400 Myr, present the characteristic N-S bar-like
structure of the young stellar content of NGC 6822. On the other
hand, the slices II, VI and IX, containing older stars with an age
of more than roughly 300 Myr, present the characteristic NE-SW
ellipsoidal spatial distribution of the old stellar content of NGC
6822. This leads to a subdivision of the stellar populations of this
galaxy into two different stellar components, according to their
spatial distribution. The first one contains stars younger than 350
$\pm$ 50 Myr, while the second one contains stars older than this
age. In Figure \ref{cmd_ellipses}, the theoretical isochrones of
300, 350 and 400 Myr illustrate this subdivision. Finally, slices
IV, VII and VIII are dominated by background stars of our Galaxy and
they do not have an adequate number of NGC 6822 stars to produce any
particular structure.

After observing the elliptical spatial distributions of slices II,
VI and IX, it is clear that there are two distinct components in the
NE and SW regions of the camera, even though old populations are
supposed to be dominated by the long-lasting effects of stellar
mixing, resulting in smooth features. This is due to the fact that
red and faint stars are expected to appear mainly in the NE and SW
regions of the camera due to variations of the PSF and the color
responses, affecting the spatial distribution, as it was mentioned
in Section 2. Nevertheless, the general direction of the old stellar
ellipse is probably unaffected, being in agreement with the Red
Giants' distribution found by Demers et al. (2006).

\section{Discussion}

\subsection{Hierarchical star formation}

The spatial distribution of the star complexes of NGC 6822, which
have been detected as described above, seem to follow The
hierarchical star formation scenario, in terms of the distribution
of star forming regions in space and their evolution in time. Star
complexes are mainly found inside larger star complexes (Figure
\ref{cmd_ellipses}). Additionally, structures younger than $\sim$ 55
Myr are clearly less extended than structures containing stars with
age up to $\sim$ 100 Myr (Figure \ref{cmd_subslices}). These
findings are in agreement with the Efremov \& Elmegreen model, where
star complexes are part of a continuous star formation hierarchy,
following the hierarchical distribution of the gas (Elmegreen \&
Efremov 1996), and star formation proceeds slower in larger star
forming regions than in smaller ones (Elmegreen \& Efremov 1996,
Efremov \& Elmegreen 1998). Similar conclusions were drawn from
previous studies of the LMC (Maragoudaki et al. 1998, Livanou et al.
2006) and the SMC (Maragoudaki et al. 2001, Livanou et al. 2007).
However, it is not clear from this study whether the first group of
detected star complexes, extended from $\sim$ 150 pc to $\sim$
300-400 pc, is characterized by lower average ages than the second
one, extended from $\sim$ 400 pc to $\sim$ 800 pc. Nevertheless, it
is clear that younger star forming regions are found to be more
concentrated than older ones.

Moreover, the presence of distinct groupings favors the existence of
hierarchical star formation, in terms of preferable sizes, in NGC
6822 and the Magellanic Clouds. There is probably consistency with
the empirically extracted hierarchy of stellar aggregates, stellar
complexes and stellar supercomplexes, within a small range in the
size limits between the first two groupings. From the lower scale
groupings to the higher, size range increases while number of star
complexes decreases. This conclusion refers to young stellar
structures and it remains to be tested also for the smallest ones
(OB associations). These structures were not found due to resolution
restriction of the observations. If the above scheme is correct and
OB associations consist another distinct group, it could be assumed
that they would span a shorter size range and they would contain
much more members. Both have been confirmed observationally in the
MCs (Livanou 2007, PhD Thesis).

Assuming that there are preferable size groupings, it is quite clear
that they cannot be detected easily from the histograms of the size
distribution of star complexes, especially in the case of the LMC.
This is not surprising, since there is usually a kind of
arbitrariness in selecting the histogram's bins and, in any case,
binning is itself a loss of information. What they can offer most is
to indicate the peak values of the size distributions. On the other
hand, diagrams with sizes sorted in ascending order together with
the Kolmogorov-Smirnov statistical tests, seem capable to visualize
groupings, stochastically study their self-consistency and reveal
their limits, providing both a qualitative and quantitative
approach. Having done this procedure, histograms of size
distribution of detected star complexes could be re-examined. For
example, the controversial LMC bimodal structure (Figure
\ref{allregions}) with its peaks centered on 550 pc and 750 pc is
probably real.

A few very large structures with size $>$ 900 pc were also found in
the MCs, three of them in the LMC and five in the SMC. Such
structures were not found in NGC 6822. Whether this is related to
some particular property of the galaxies, it is not clear from this
investigation.

\subsection{A possible strong interaction}

The examination of how the different stellar populations are
distributed in space reveals two different structures: The N-S
bar-like structure of young stars and the NE-SW elliptical structure
of the old stars. The non similar spatial distribution of young and
old stars has been pointed out by previous studies (Hodge 1977; de
Blok, \& Walter 2003; Komiyama et al. 2003; Battinelli et al. 2003;
Demers et al. 2006). What is not commented is the unlikelihood of a
smooth transition from the younger structure to the older one by
stellar dissemination. In a quiescent and undisturbed galaxy, one
would expect that the stars, after their creation, would drift
and/or participate to the expansion of the OB associations in which
they were born. This would probably result in an old population
embracing the observed star forming regions. In the case of NGC
6822, star formation is taking place in a large scale, along the
North-South direction. The extended old stellar component, however,
does not embrace the star forming regions, but it has a rather
different alignment. This probably means that the huge young
structure is too young to have already produced an old 'halo' around
it. This very recent and spatially extended star formation, together
with the presence of the highly perturbed gas disk, bring up the
possibility of a past strong event (Demers et al. 2006). Given the
lack of a nearby massive galaxy to NGC 6822 which could tidally
disrupt it after one or more close encounters, a merging event could
be a reasonable scenario. Such an event could be responsible for the
excess of the recent star formation and the asymmetric gas disk.

Whatever its nature, a strong interaction could have a lot of
consequences to what has been concluded up to nowadays for NGC 6822.
The interaction scenario between NGC 6822 and the NW Cloud (de Blok,
\& Walter 2000, 2006) could be insufficient to explain such large
differences between the young and the old stellar populations. The
mass of the Cloud is $\sim 1.4 \times 10^{7} M_{\sun}$ (de Blok, \&
Walter 2000), only 5 \% of the total observed baryonic mass of NGC
6822 ($\sim 2.8 \times 10^{8} M_{\sun}$, Weldrake et al. 2003). The
suggestion that the Cloud is a separated galaxy has to be treated
with caution. One of the main arguments supporting this is that the
NW half of atomic hydrogen contains 20\% more mass than the SE half,
as measured with respect to a minor axis passing through the
geometrical center. This was based on the assumption of an
intrinsically symmetric gas disk (de Blok, \& Walter 2000). However,
from the same work it is revealed for the first time that the gas
disk of NGC 6822 is highly perturbed. Additionally, if NGC 6822 has
undergone a strong interaction in the past, the assumption of a
symmetric gas disc could be questionable. Whether this gas Cloud is
a separated galaxy needs further investigation. The indication of a
strong past interaction also strengthens the argument of NGC 6822
being a Polar Ring galaxy (Demers et al. 2006), as it provides the
most important prerequisite for its formation.

The young large scale structure of NGC 6822 (\ref{9regions}) is
associated with stars younger than 350 $\pm$ 50 Myr, while the old
large scale structure (which is misaligned with respect to the young
one) is associated with stars older than 350 $\pm$ 50 Myr, dating
the possible strong interaction before that time. This timescale is
in a satisfactory agreement with the timescales of the increase of
the recent SFR and the formation of the tidal gas features (see
Introduction). Furthermore, Demers et al. (2006) place the strong
interaction necessary for the formation of the Polar Ring well
before 500 Myr. The validity of the 350 $\pm$ 50 Myr timescale is
also confirmed by the fact that the stars of NGC 6822 probably need
400 Myr or even 500-600 Myr to mix (Wyder 2001). Thus, the time
limit between the young and the old large scale structures is
probably unbiased by stellar mixing.

Although from the present study there are indications that NGC 6822
could have undergone a strong and not a weak interaction in the
past, it is not possible to derive certain conclusions from our
data. Further investigation is needed towards a more definite
answer.

\section{Conclusions}

We used the "Local Group Survey" stellar catalogue of NGC 6822 to
identify its star complexes. These regions were detected from the
isopleths, based on counts of the young stars, above a statistical
cutoff limit, resulting in a list of the positions and sizes of star
complexes.

Indications of hierarchical star formation were found in NGC 6822,
in terms of spatial distribution and time evolution. Star complexes
are mainly found inside larger star complexes, and younger star
forming structures are clearly less extended than the older ones.
These findings add support to the Efremov \& Elmegreen model
(Elmegreen \& Efremov 1996, Efremov \& Elmegreen 1998), where star
complexes are part of a continuous star formation hierarchy,
following the hierarchical distribution of the gas, and star
formation proceeds slower in larger star forming regions than in
smaller ones. Additionaly, indications of hierarchical star
formation in terms of preferable sizes of the star complexes were
found in NGC 6822 and  the Magellanic Clouds. The diagrams of the
sizes of all the detected star complexes of NGC 6822 and the
Magellanic Clouds, sorted in ascending order, were used along with
the two-tailed Kolmogorov-Smirnov statistical test to search for
different size groups: Two main groups were identified, the first
one ranging from $\sim$ 150 pc to $\sim$ 300-400 pc and the second
up to $\sim$ 800 pc. This is probably consistent with the
empirically extracted hierarchy of stellar aggregates, stellar
complexes and stellar supercomplexes. The smallest size groups, like
OB associations, remain to be tested whether they form a distinct
group.

The large scale structures of the NGC 6822 galaxy were studied by
selecting stellar populations of various ages from the
Colour-Magnitude Diagram and projecting their stellar content in
space. It was found that the old population does not contain the
young one, but rather they form two distinct structures rotated in
position angle. A possible strong interaction could explain this
misalignment and, given that NGC 6822 is relatively isolated in
space, a merging event could be considered as a reasonable scenario.
The traces of the possible strong interaction on the spatial
distribution of the stellar populations are dating before 350 $\pm$
50 Myr, being in agreement with the recent star formation excess of
this galaxy.

\begin{acknowledgements}
      We are grateful to the anonymous referee for comments and
      suggestions that significally improved this paper.
      This work was supported by the Special Account for
      Research Grants of the National and Kapodistrian University of
      Athens. We would like to thank the Local Group Survey Team for
      publicly releasing their data. We are also grateful to C. Gallart (Instituto de
      Astrof\'{\i}sica de Canarias, La Laguna, Spain) who kindly offered her
      data for tests.
\end{acknowledgements}


\end{document}